
\input harvmac

\noblackbox
\baselineskip 20pt plus 2pt minus 2pt

\overfullrule=0pt



\def\bs{\bigskip}
\def\no{\noindent}
\def\hb{\hfill\break}
\def\qq{\qquad}
\def\bl{\bigl}
\def\br{\bigr}

\def\IR{\relax{\rm I\kern-.18em R}}

\def\np {  Nucl. Phys. }


\def\r{\rho}
\def\a{\alpha}

\def\b{\beta}

\def\g{\gamma}
\def\G{\Gamma}
\def\d{\delta}

\def\e{\epsilon}

\def\th{\theta}

\def\m{\mu}
\def\n{\nu}
\def\om{\omega}
\def\Om{\Omega}
\def\l{\lambda}
\def\L{\Lambda}
\def\s{\sigma} 
\def\S{\Sigma}
\def\vphi{\varphi}

\def\IR{\relax{\rm I\kern-.18em R}}

\def \ha {{1\over 2}}

\def \ov {\over}

\def\const{{\rm const.}}



\lref\BSthree{I. Bars and K. Sfetsos, Mod. Phys. Lett. {\bf A7} (1992) 1091.}

\lref\BShet{I. Bars and K. Sfetsos, Phys. Lett. {\bf 277B} (1992) 269.}

\lref\BSglo{I. Bars and K. Sfetsos, Phys. Rev. {\bf D46} (1992) 4495.}

\lref\BSexa{I. Bars and K. Sfetsos, Phys. Rev. {\bf D46} (1992) 4510.} 

\lref\SFET{K. Sfetsos, Nucl. Phys. {\bf B389} (1993) 424.}

\lref\BSslsu{I. Bars and K. Sfetsos, Phys. Lett. {\bf 301B} (1993) 183.}

\lref\BSeaction{I. Bars and K. Sfetsos, Phys. Rev. {\bf D48} (1993) 844.}

\lref\BN{ I. Bars and D. Nemeschansky, Nucl. Phys. {\bf B348} (1991) 89.}

\lref\BCR{K. Bardakci, M. Crescimanno and E. Rabinovici, 
Nucl. Phys. {\bf B344} (1990) 344.}

\lref\WIT{E. Witten, Phys. Rev. {\bf D44} (1991) 314.}

 \lref\IBhet{ I. Bars, Nucl. Phys. {\bf B334} (1990) 125. }

 \lref\IBCS{ I. Bars, ``String Propagation on Black Holes'', USC-91-HEP-B3.\hb
{\it Curved Space-time Strings and Black Holes}, 
in Proc. 
 {\it XX$^{th}$ Int. Conf. on Diff. Geometrical Methods in Physics}, eds. S. 
 Catto and A. Rocha, Vol. 2, p. 695, (World Scientific, 1992).}

 \lref\CRE{M. Crescimanno, Mod. Phys. Lett. {\bf A7} (1992) 489.}

\lref\MSW{G. Mandal, A. Sengupta and S. Wadia, 
Mod. Phys. Lett. {\bf A6} (1991) 1685.}

 \lref\HOHO{J.B. Horne and G.T. Horowitz, Nucl. Phys. {\bf B368} (1992) 444.}

 \lref\FRA{E. S. Fradkin and V. Ya. Linetsky, Phys. Lett. {\bf 277B} 
          (1992) 73.}

 \lref\ISH{N. Ishibashi, M. Li, and A. R. Steif, 
         Phys. Rev. Lett. {\bf 67} (1991) 3336.}

 \lref\HOR{P. Horava, Phys. Lett. {\bf 278B} (1992) 101.}

 \lref\RAI{E. Raiten, ``Perturbations of a Stringy Black Hole'',
         Fermilab-Pub 91-338-T.}

 \lref\GER{D. Gershon, Phys. Rev. {\bf D49} (1994) 999.}

\lref\GERexa{D. Gershon, ``Semiclassical vs. Exact Solutions of charged Black 
Hole in four dimensions and exact $O(D,D)$ duality'', TAUP-2121-93, 
hepth/9311122.}

\lref\GERexadual{D. Gershon, ``Exact $O(D,D)$ transformations in WZW models,
TAUP-2129-93, hepth/9312154.}

 \lref \GIN {P. Ginsparg and F. Quevedo,  Nucl. Phys. {\bf B385} (1992) 527. }

 \lref\HOHOS{ J.H. Horne, G.T. Horowitz and A. R. Steif, Phys. Rev. Lett.
 {\bf 68} (1991) 568.}

 \lref\groups{
 M. Crescimanno. Mod. Phys. Lett. {\bf A7} (1992) 489. \hb
 J. B. Horne and G.T. Horowitz, Nucl. Phys. {\bf B368} (1992) 444. \hb
 E. S. Fradkin and V. Ya. Linetsky, Phys. Lett. {\bf 277B} (1992) 73. \hb
 P. Horava, Phys. Lett. {\bf 278B} (1992) 101.\hb
 E. Raiten, ``Perturbations of a Stringy Black Hole'',
         Fermilab-Pub 91-338-T.\hb
 D. Gershon, ``Exact Solutions of Four-Dimensional Black Holes in 
         String Theory'', TAUP-1937-91.}

\lref\NAWIT{C. Nappi and E. Witten, Phys. Lett. {\bf 293B} (1992) 309.}

\lref\FRATSE{E. S. Fradkin and A.A. Tseytlin, 
Phys. Lett. {\bf 158B} (1985) 316.}

\lref\DASA{ S. Das and B. Sathiapalan, Phys. Rev. Lett. {\bf 56} (1986) 2664.}

\lref\CALLAN{ C.G. Callan, D. Friedan, E.J. Martinec and M. Perry, 
Nucl. Phys. {\bf B262} (1985) 593.}

\lref\DB{L. Dixon, J. Lykken and M. Peskin, Nucl. Phys.
{\bf B325} (1989) 325.}

\lref\IB{I. Bars, Nucl. Phys. {\bf B334} (1990) 125.}

\lref\BUSCHER{T. Buscher, Phys. Lett. {\bf 194B} (1987) 59;
Phys. Lett. {\bf 201B} (1988) 466.}

\lref\RV{M. Rocek and E. Verlinde, Nucl. Phys. {\bf B373} (1992) 630.}
\lref\GR{A. Giveon and M. Rocek, Nucl. Phys. {\bf B380} (1992) 128. }

\lref\nadual{X. C. de la Ossa and F. Quevedo, 
Nucl. Phys. {\bf B403} (1993) 377.}

\lref\duearl{B.E. Fridling and A. Jevicki, 
Phys. lett. {\bf B134} (1984) 70.\hb
E.S. Fradkin and A.A. Tseytlin, Ann. Phys. {\bf 162} (1985) 31.}

\lref\ABEN{I. Antoniadis, C. Bachas, J. Ellis and D.V. Nanopoulos,
Phys. Lett. {\bf B211} (1988) 393.}

\lref\Kalofour{N. Kaloper, Phys. Rev. {\bf D48} (1993) 4658.}

\lref\GPS{S.B. Giddings, J. Polchinski and A. Strominger,
Phys. Rev. {\bf D48} (1993) 5784.}

\lref\SENrev{A. Sen, ``Black Holes and Solitons in String Theory'', 
TIFR-TH-92-57.}

\lref\TSEd{A.A. Tseytlin, Mod. Phys. Lett. {\bf A6} (1991) 1721.}

\lref\TSESC{A. S. Schwarz and A.A. Tseytlin, ``Dilaton shift under duality
and torsion of elliptic complex'', IMPERIAL/TP/92-93/01. }

\lref\Dualone{K. Meissner and G. Veneziano, 
Phys. Lett. {\bf B267} (1991) 33;
Mod. Phys. Lett. {\bf A6} (1991) 3397. \hb
M. Gasperini and G. Veneziano, Phys. Lett. {\bf 277B} (1992) 256. \hb
M. Gasperini, J. Maharana and G. Veneziano, Phys. Lett. {\bf 296B} (1992) 51.}

\lref\rovr{K. Kikkawa and M. Yamasaki, Phys. Lett. {\bf B149} (1984) 357.\hb
N. Sakai and I. Senda, Prog. theor. Phys. {\bf 75} (1986) 692.}

\lref\narain{K.S. Narain, Phys. Lett. {\bf B169} (1986) 369.\hb
K.S. Narain, M.H. Sarmadi and C. Vafa, Nucl. Phys. {\bf B288} (1987) 551.}

\lref\GV{P. Ginsparg and C. Vafa, Nucl. Phys. {\bf B289} (1987) 414.}
\lref\nssw{V. Nair, A. Shapere, A. Strominger and F. Wilczek, 
Nucl. Phys. {\bf B287} (1987) 402.}

\lref\vafa{C. Vafa, ``Strings and Singularities'', HUTP-93/A028.}

\lref\Dualtwo{A. Sen, 
Phys. Lett. {\bf B271} (1991) 295;\ ibid. {\bf B274} (1992) 34;
Phys. Rev. Lett. {\bf 69} (1992) 1006. \hb
S. Hassan and A. Sen, Nucl. Phys. {\bf B375} (1992) 103. \hb
J. Maharana and J. H. Schwarz, Nucl. Phys. {\bf B390} (1993) 3.\hb
A. Kumar, Phys. Lett. {\bf B293} (1992) 49.}

\lref\dualmargi{S.F. Hassan and A. Sen, Nucl. Phys. {\bf B405} (1993) 143;
M. Henningson and C. Nappi, Phys. Rev. {\bf D48} (1993) 861;
E. Kiritsis, Nucl. Phys. {\bf B405} (1993) 109.}

\lref\bv{R. Brandenberger and C. Vafa, Nucl. Phys. {\bf B316} (1989) 301.}
\lref\gsvy{B.R. Greene, A. Shapere, C. Vafa and S.T. Yau, 
Nucl. Phys. {\bf B337} (1990) 1.}
\lref\tv{A.A. Tseytlin and C. Vafa,  Nucl. Phys. {\bf B372} (1992) 443.}

\lref\grvm{A. Shapere and F. Wilczek, Nucl. Phys. {\bf B320} (1989) 609.\hb
A. Giveon, E. Rabinovici and G. Veneziano, 
Nucl. Phys. {\bf B322} (1989) 167.\hb
A. Giveon, N. Malkin and E. Rabinovici, Phys. Lett. {\bf B238} (1990) 57.}

\lref\GRV{M. Gasperini, R. Ricci and G. Veneziano,
Phys. Lett. {\bf B319} (1993) 438.}

\lref\KIRd{E. Kiritsis, Nucl. Phys. {\bf B405} (1993) 109.}

\lref\dualrev{For reviews see, 
A. Giveon, M. Porrati and E. Rabinovici, Phys. Rep. {\bf 244} (1994)
77; E. Alvarez, L. Alvarez-Gaume and Y. Lozano, Nucl. Phys. (Proc. supp.)
{\bf 41} (1995) 1.}


\lref\slt{A. Giveon, Mod. Phys. Lett. {\bf A6} (1991) 2843.}

\lref\GIPA{A. Giveon and A. Pasquinucci, ``On cosmological string backgrounds 
with toroidal isometries'', IASSNS-HEP-92/55, August 1992.}

\lref\KASU{Y. Kazama and H. Suzuki, Nucl. Phys. {\bf B234} (1989) 232. \hb
Y. Kazama and H. Suzuki Phys. Lett. {\bf 216B} (1989) 112.}

\lref\WITanom{E. Witten, Comm. Math. Phys. {\bf 144} (1992) 189.} 

\lref\WITnm{E. Witten, Nucl. Phys. {\bf B371} (1992) 191.}

\lref\IBhetero{I. Bars, Phys. Lett. {\bf 293B} (1992) 315.}

\lref\IBerice{I. Bars, {\it Superstrings on Curved Space-times}, Lecture
delivered at the Int. workshop on {\it String Quantum Gravity and Physics
at the Planck Scale}, Erice, Italy, June 1992.}

\lref\DVV{R. Dijkgraaf, E. Verlinde and H. Verlinde, Nucl. Phys. {\bf B371}
(1992) 269.}
 
\lref\TSEY{A.A. Tseytlin, Phys. Lett. {\bf 268B} (1991) 175.}

\lref\JJP{I. Jack, D. R. T. Jones and J. Panvel, 
          Nucl. Phys. {\bf B393} (1993) 95.} 

\lref\BST { I. Bars, K. Sfetsos and A.A. Tseytlin, unpublished. }

\lref\TSEYT{ A.A. Tseytlin, Nucl. Phys. {\bf B399} (1993) 601.}

\lref\TSEYTt{A.A. Tseytlin, Nucl. Phys. {\bf B411} (1993) 509.}

 \lref\SHIF { M. A. Shifman, Nucl. Phys. {\bf B352} (1991) 87.}
\lref\SHIFM { H. Leutwyler and M. A. Shifman, Int. J. Mod. Phys. {\bf
A7} (1992) 795. } 

\lref\POLWIG { A. M. Polyakov and P. B. Wiegman, Phys.
Lett. {\bf 141B} (1984) 223.  } 

\lref\BCR{K. Bardakci, M. Crescimanno
and E. Rabinovici, Nucl. Phys. {\bf B344} (1990) 344. }

\lref\Wwzw{E. Witten, Commun. Math. Phys. {\bf 92} (1984) 455.}

\lref\GKO{P. Goddard, A. Kent and D. Olive, Phys. Lett. {\bf 152B} (1985) 88.}

\lref\Toda{A. N. Leznov and M. V. Saveliev, Lett. Math. Phys. {\bf 3} (1979) 
489. \hb A. N. Leznov and M. V. Saveliev, Comm. Math. Phys. {\bf 74}
(1980) 111.}

\lref\GToda{J. Balog, L. Feh\'er, L. O'Raifeartaigh, P. Forg\'acs and A. Wipf,
Ann. Phys. (New York) {\bf 203} (1990) 76; Phys. Lett. {\bf 244B}
(1990) 435.}

\lref\GWZW{ E. Witten, \np {\bf B223} (1983) 422. \hb
K. Bardakci, E. Rabinovici and B. S\"aring, Nucl. Phys. {\bf B299}
(1988) 157. \hb K. Gawedzki and A. Kupiainen, Phys. Lett. {\bf 215B}
(1988) 119.; Nucl. Phys. {\bf B320} (1989) 625. }

\lref\SCH{ D. Karabali, Q-Han Park, H. J. Schnitzer and Z. Yang, 
                   Phys. Lett. {\bf B216} (1989) 307. \hb D. Karabali
and H. J. Schnitzer, Nucl. Phys. {\bf B329} (1990) 649. }

 \lref\KIR{E. Kiritsis, Mod. Phys. Lett. {\bf A6} (1991) 2871. }

\lref\BIR{N. D. Birrell and P. C. W. Davies, 
{\it Quantum Fields in Curved Space}, Cambridge University Press.}

\lref\WYB{B. G. Wybourn, {\it Classical Groups for Physicists }
(John Wiley \& sons, 1974).}

\lref\Brinkman{H.W. Brinkmann, Math. Ann. {\bf 94} (1925) 119.}

\lref\SANTA{R. Guven, Phys. Lett. {\bf 191B} (1987) 275.\hb
D. Amati and C. Klimcik, Phys. Lett. {\bf 219B} (1989) 443.\hb
G.T. Horowitz and A. R. Steif, Phys. Rev. Lett. {\bf 64} (1990) 260;
Phys. Rev. {\bf D42} (1990) 1950.\hb
R.E. Rudd, Nucl. Phys. {\bf B352} (1991) 489.\hb
C. Duval, G.W. Gibbons and P.A. Horvathy, Phys. Rev. {\bf D43} (1991) 3907.\hb
C. Duval, Z. Horvath and P.A. Horvathy, Phys. Lett. {\bf B313} (1993) 10.\hb
E.A. Bergshoeff, R. Kallosh and T. Ortin, Phys. Rev. {\bf D47} (1993) 5444.}  
\lref\SANT{J. H. Horne, G.T. Horowitz and A. R. Steif, 
Phys. Rev. Lett. {\bf 68} (1991) 568.}

\lref\tsecov{A.A. Tseytlin, Nucl. Phys. {\bf B390} (1993) 153; 
Phys. Rev. {\bf D47} (1993) 3421.}

\lref\garriga{J. Garriga and E. Vardaguer, Phys. Rev. {\bf D43} (1991) 391.}

\lref\PRE{J. Prescill, P. Schwarz, A. Shapere, S. Trivedi and F. Wilczek, 
Mod. Phys Lett. {\bf A6} (1991) 2353.\hb
C. Holzhey and F. Wilczek, Nucl. Phys. {\bf B380} (1992) 447.}

\lref\HAWK{J. B. Hartle and S. W. Hawking Phys. Rev. {\bf D13} (1976) 2188.\hb
S. W. Hawking, Phys. Rev. {\bf D18} (1978) 1747.}

\lref\HAWKI{S. W. Hawking, Comm. Math. Phys. {\bf 43} (1975) 199.}

\lref\HAWKII{S. W. Hawking, Phys. Rev. {\bf D14} (1976) 2460.}

\lref\euclidean{S. Elitzur, A. Forge and E. Rabinovici,
Nucl. Phys. {\bf B359} (1991) 581. }

\lref\ITZ{C. Itzykson and J. Zuber, {\it Quantum Field Theory}, 
McGraw Hill (1980). }

\lref\kacrev{P. Goddard and D. Olive, Journal of Mod. Phys. {\bf A} Vol. 1,
No. 2 (1986) 303.}

\lref\BBS{F.A. Bais, P. Bouwknegt, K.S. Schoutens and M. Surridge, 
Nucl. Phys. {\bf B304} (1988) 348.}

\lref\nonl{A. Polyakov, {\it Fields, Strings and Critical Phenomena}, Proc. of
Les Houses 1988, eds. E. Brezin and J. Zinn-Justin North-Holland, 1990.\hb
Al. B. Zamolodchikov, preprint ITEP 87-89. \hb
K. Schoutens, A. Sevrin and P. van Nieuwenhuizen, Proc. of the Stony Brook 
Conference {\it Strings and Symmetries 1991}, World Scientific, 
Singapore, 1992. \hb
J. de Boer and J. Goeree, ``The Effective Action of $W_3$ Gravity to all
\hb orders'', THU-92/33.}

\lref\HOrev{G.T. Horowitz, {\it The Dark Side of String Theory:
Black Holes and Black Strings}, Proc. of the 1992 Trieste Spring School on
String Theory and Quantum Gravity.}

\lref\HSrev{J. Harvey and A. Strominger, {\it Quantum Aspects of Black 
Holes}, Proc. of the 1992 Trieste Spring School on
String Theory and Quantum Gravity.}

\lref\GM{G. Gibbons, Nucl. Phys. {\bf B207} (1982) 337.\hb
G. Gibbons and K. Maeda, Nucl. Phys. {\bf B298} (1988) 741.}

\lref\GID{S. B. Giddings, Phys. Rev. {\bf D46} (1992) 1347.}

\lref\PRErev{J. Preskill, {\it Do Black Holes Destroy Information?},
Proc. of the International Symposium on Black Holes, Membranes, Wormholes,
and Superstrings, The Woodlands, Texas, 16-18 January, 1992.}

\lref\tye{S-W. Chung and S. H. H. Tye, Phys. Rev. {\bf D47} (1993) 4546.}

\lref\eguchi{T. Eguchi, Mod. Phys. Lett. {\bf A7} (1992) 85.}

\lref\blau{M. Blau and G. Thompson, Nucl. Phys. {\bf B408} (1993) 345.}

\lref\HSBW{P. S. Howe and G. Sierra, Phys. Lett. {\bf 144B} (1984) 451.\hb
J. Bagger and E. Witten, Nucl. Phys. {\bf B222} (1983) 1.}

\lref\GSW{M. B. Green, J. H. Schwarz and E. Witten, {\it Superstring Theory},
Cambridge Univ. Press, Vols. 1 and 2, London and New York (1987).}

\lref\KAKU{M. Kaku, {\it Introduction to Superstrings}, Springer-Verlag, Berlin
and New York (1991).}

\lref\LSW{W. Lerche, A. N. Schellekens and N. P. Warner, {\it Lattices and
Strings }, Physics Reports {\bf 177}, Nos. 1 \& 2 (1989) 1, North-Holland,
Amsterdam.}

\lref\confrev{P. Ginsparg and J. L. Gardy in {\it Fields, Strings, and
Critical Phenomena}, 1988 Les Houches School, E. Brezin and J. Zinn-Justin, 
eds, Elsevier Science Publ., Amsterdam (1989). \hb
J. Bagger, {\it Basic Conformal Field Theory},
Lectures given at 1988 Banff Summer Inst. on Particle and Fields,
Banff, Canada, Aug. 14-27, 1988, HUTP-89/A006, January 1989. }

\lref\CHAN{S. Chandrasekhar, {\it The Mathematical Theory of Black Holes},
Oxford University Press, 1983.}

\lref\KOULU{C. Kounnas and D. L\"ust, Phys. Lett. {\bf 289B} (1992) 56.}

\lref\PERRY{M. J. Perry and E. Teo, Phys. Rev. Lett. {\bf 70} (1993) 2669.\hb
P. Yi, Phys. Rev. {\bf D48} (1993) 2777.}

\lref\GiKi{A. Giveon and E. Kiritsis, Nucl. Phys. {\bf B411} (1994) 487.}

\lref\kar{S.K. Kar and A. Kumar, Phys. Lett. {\bf 291B} (1992) 246.}

\lref\NW{C. Nappi and E. Witten, Phys. Rev. Lett. {\bf 71} (1993) 3751.}

\lref\HK{M. B. Halpern and E. Kiritsis, 
Mod. Phys. Lett. {\bf A4} (1989) 1373.}

\lref\MOR{A.Yu. Morozov, A.M. Perelomov, A.A. Rosly, M.A. Shifman and 
A.V. Turbiner, Int. J. Mod. Phys. {\bf A5} (1990) 803.}

\lref\KK{E. Kiritsis and C. Kounnas, Phys. Lett. {\bf B320} (1994) 264.}

\lref\GSVYdyn{ B.R. Greene, A. Shapere, C. Vafa, S.-T. Yau, 
Nucl. Phys. {\bf B337} (1990) 1.}
 
\lref\KKdyn{E. Kiritsis and C. Kounnas, Phys. Lett. {\bf B331} (1994) 51.}

\lref\Tsdyn{A.A. Tseytlin, Phys. Lett. {\bf B334} (1994) 315.}

\lref\KST{K. Sfetsos and A.A. Tseytlin, Phys. Rev. {\bf D49} (1994) 2933.}

\lref\KSTh{K. Sfetsos and A.A. Tseytlin, Nucl. Phys. {\bf B415} (1994) 116.}

\lref\KP{S.P. Khastgir and A. Kumar, ``Singular limits and string solutions'',
IP/BBSR/93-72, hepth/9311048.}

\lref\etc{K. Sfetsos, Phys. Lett. {\bf B324} (1994) 335.}

\lref\KTone{C. Klimcik and A.A. Tseytlin, Phys. Lett. {\bf B323} (1994) 305.}

\lref\KTtwo{C. Klimcik and A.A. Tseytlin, ``Exact four dimensional string
solutions and Toda-like sigma models from `null-gauged' WZNW theories,
Imperial/TP/93-94/17, PRA-HEP 94/1, hepth/9402120.}

\lref\saletan{E.J. Saletan, J. Math. Phys. {\bf 2} (1961) 1.}
\lref\jao{D. Cangemi and R. Jackiw, Phys. Rev. Lett. {\bf 69} (1992) 233.}
\lref\jat{D. Cangemi and R. Jackiw, Ann. Phys. (NY) {\bf 225} (1993) 229.}

\lref\ORS{ D. I. Olive, E. Rabinovici and A. Schwimmer, Phys. Lett. {\bf B321}
(1994) 361.}

\lref\edc{K. Sfetsos, ``Exact String Backgrounds from WZW models based on 
Non-semi-simple groups'', THU-93/31, hepth/9311093, to appear in 
Int. J. Mod. Phys. {\bf A} (1994).}

\lref\sfedual{K. Sfetsos, Phys. Rev. {\bf D50} (1994) 2784.}

\lref\grnonab{A. Giveon and M. Rocek, Nucl. Phys. {\bf B421} (1994) 173.}

\lref\wigner{E. ${\rm In\ddot on\ddot u}$ and E.P. Wigner, 
Proc. Natl. Acad. Sci. U. S. {\bf 39} (1953) 510.}

\lref\HY{ J. Yamron and M.B. Halpern, Nucl. Phys. {\bf B351} (1991) 333.}
\lref\HB{K. Bardakci and M.B. Halpern, Phys. Rev. {\bf D3} (1971) 2493.}
\lref\MBH{M.B. Halpern, Phys. Rev {\bf D4} (1971) 2398.}
\lref\balog{J. Balog, L. O'Raifeartaigh, P. Forgacs and A. Wipf,
Nucl. Phys. {\bf B325} (1989) 225.}
\lref\kacm{V. G. Kac, Funct. Appl. {\bf 1} (1967) 328.\hb
R.V. Moody, Bull. Am. Math. Soc. {\bf 73} (1967) 217.}

\lref\TSEma{A.A. Tseytlin, Nucl. Phys. {\bf B418} (1994) 173.}

\lref\BCHma{J. de Boer, K. Clubock and M.B. Halpern, ``Linearized form
of the Generic Affine-Virasoro Action'', UCB-PTH-93/34, hepth/9312094.}

\lref\AABL{E. Alvarez, L.Alvarez-Gaume, J.L.F. Barbon and Y. Lozano,
Nucl. Phys. {\bf B415} (1994) 71.}

\lref\noumo{N. Mohammedi,  Phys. Lett. {\bf B325} (1994) 371.}

\lref\sfetse{K. Sfetsos and A.A. Tseytlin, unpublished (February 1994).}
\lref\figsonia{J.M. Figueroa-O'Farrill and S. Stanciu, Phys. Lett. {\bf B327} (\1994) 40.} 

\lref\ALLleadord{E. Witten, Phys. Rev. {\bf D44} (1991) 314.\hb
J.B. Horne and G.T. Horowitz, Nucl. Phys. {\bf B368} (1992) 444.\hb
M. Crescimanno, Mod. Phys. Lett. {\bf A7} (1992) 489.\hb
I. Bars and K. Sfetsos, Mod. Phys. Lett. {\bf A7} (1992) 1091.\hb
E.S. Fradkin and V.Ya. Linetsky, Phys. Lett. {\bf 277B} (1992) 73. \hb
I. Bars and K. Sfetsos, Phys. Lett. {\bf 277B} (1992) 269.\hb
P. Horava, Phys. Lett. {\bf 278B} (1992) 101.\hb
E. Raiten, ``Perturbations of a Stringy Black Hole'', Fermilab-Pub 91-338-T.\hb
P. Ginsparg and F. Quevedo,  Nucl. Phys. {\bf B385} (1992) 527.\hb
more}

\lref\ALLexact{
R. Dijkgraaf, E. Verlinde and H. Verlinde, Nucl. Phys. {\bf B371} (1992) 269.
\hb
I. Bars and K. Sfetsos, Phys. Rev. {\bf D46} (1992) 4510.;
Phys. Lett. {\bf 301B} (1993) 183.\hb
K. Sfetsos, Nucl. Phys. {\bf B389} (1993) 424.\hb
Gerhon}

\lref\DHooft{T. Dray and G. 't Hooft, Nucl. Phys. {\bf B253} (1985)
173.} 
\lref\MyPe{ R.C. Myers and M.J. Perry, Ann. of Phys. {\bf 172} (1986)
304.} 
\lref\GiMa{G. Gibbons and K. Maeda, Nucl. Phys. {\bf B298} (1988) 741.}
\lref\KLOPV{R. Kallosh, A. Linde, T. Ort\'\i n, A. Peet and A. Van Proeyen,
Phys. Rev. {\bf D46} (1992) 5278.}

\lref\lousan{ C.O. Loust\'o and N. S\'anchez, Int. J. Mod. Phys. {\bf A5}
(1990) 915; Nucl. Phys. {\bf B355} (1991) 231.\hb
V. Ferrari and P. Pendenza, Gen. Rel. Grav. {\bf 22} (1990)
1105.\hb 
H. Balasin and H. Nachbagauer, Class. Quantum Grav. {\bf 12} (1995) 707. \hb
K. Hayashi and T. Samura, Phys. Rev. {\bf D50} (1994) 3666. \hb
S. Das and P. Majumdar, Phys. Lett. {\bf B348} (1995) 349.}


\lref\lousanII{ C.O. Loust\'o and N. S\'anchez, ``Scattering processes at the 
Planck scale'', UAB-FT-353, gr-qc/9410041 and references therein.}

\lref\DHooftII{ T. Dray and G. 't Hooft, Commun. Math. Phys. {\bf 99}
(1985) 613.}
\lref\DHooftIII{T. Dray and G. 't Hooft,
Class. Quant. Grav. {\bf 3} (1986) 825.}

\lref\DHooftII{ T. Dray and G. 't Hooft, Commun. Math. Phys. {\bf 99}
(1985) 613; Class. Quant. Grav. {\bf 3} (1986) 825.}

\lref\VV{H. Verlinde and E. Verlinde, Nucl. Phys. {\bf B371} (1992)
246.}

\lref\aisexl{P.C. Aichelburg and R.U. Sexl, Gen. Rel. Grav. {\bf 2}
(1971) 303.}
\lref\cagan{C.G. Callan and Z. Gan, Nucl. Phys. {\bf B272} (1986) 647.}
\lref\hooft{ G. 't Hooft, Nucl. Phys. {\bf B335} (1990) 138.}
\lref\hoo{G. 't Hooft, Phys. Lett. {\bf B198} (1987) 61;
Nucl. Phys. {\bf B304} (1988) 867.}

\lref\tasepr{ I.S. Gradshteyn and I.M. Ryznik, {\it Tables of integrals,
Series and Products}, Academic, New York, (1980). }
\lref\HOTA{ M. Hotta and M. Tanaka, Class. Quantum Grav. {\bf 10} (1993) 307.}
\lref\otth{ V. Ferrari and P. Pendenza, Gen. Rel. Grav. {\bf 22} (1990)
1105.\hb 
H. Balasin and H. Nachbagauer, Class. Quantum Grav. {\bf 12} (1995) 707. \hb
K. Hayashi and T. Samura, Phys. Rev. {\bf D50} (1994) 3666.}

\lref\KSHM{ D. Kramer, H. Stephani, E. Herlt and M. MacCallum,
{\it Exact Solutions of Einstein's Field Equations}, Cambridge (1980).}

\lref\lathos{ C.O. Loust\'o and N. S\'anchez, Phys. Lett. {\bf B220} (1989)
55.} 

\lref\BAIT{ M. Bander and C. Itzykson, Rev. of Mod. Phys. {\bf 38} (1966) 330;
Rev. of Mod. Phys. {\bf 38} (1966) 346.}

\lref\rpen{R. Penrose, in General Relativity: papers in honour of J.L. Synge,
ed. L. O'Raifeartaigh (Clarendon, Oxford, 1972) 101.}
\lref\AGV{L.N. Lipatov, Nucl. Phys. {\bf B365} (1991) 614.\hb
R. Jackiw, D. Kabat and M. Ortiz,
Phys. Lett. {\bf B277} (1992) 148.\hb
D. Kabat and M. Ortiz, Nucl. Phys. {\bf B388} (1992) 570.\hb
D. Amati, M. Ciafaloni and G. Veneziano,
Nucl. Phys. {\bf B403} (1993) 707.}


\lref\KPen{ K.A. Khan and R. Penrose, Nature (London) {\bf 229} (1971) 185.}
\lref\DOVE{M. Dorca and E. Verdaguer, Nucl. Phys. {\bf B403} (1993) 770.}

\lref\xanth{S. Cahndrasekhar and B. Xanthopoulos }

\lref\stplane{K. Sfetsos and A.A. Tseytlin, Nucl. Phys. {\bf B427} (1994) 245.}

\lref\shock{ K. Sfetsos, Nucl. Phys. {\bf B436} (1995) 721.}

\lref\gardiner{C.W. Gardiner, {\it Handbook of stochastic methods for Physics, 
Chemistry and the Natural sciences}, Berlin, Springer, 1983.}

\lref\CHSW{ S. Chaudhuri and J.A. Schwarz, Phys. Lett. {\bf B219} (1989) 291.}
 
\lref\DJHOT{ D.H. Hartley, M. \"Onder and R.W. Tucker, 
Class. Quantum Grav. {\bf 6} (1989) 1301\hb
T. Dray and P. Joshi, Class. Quantum Grav. {\bf 7} (1990) 41.}

\lref\CHS{ C.G. Callan, J.A. Harvey and A. Strominger, Nucl. Phys. {\bf B359}
(1991) 611.}

\lref\schtalk{ J.H. Schwarz, 
``Evidence for Non-perturbative String Symmetries'',
CALT-68-1965, hepth/9411178.}

\lref\bakasII{I. Bakas, Phys. Lett. {\bf B343} (1995) 103.}
\lref\basfep{I. Bakas and K. Sfetsos, Phys. Lett. {\bf B349} (1995) 448.}
\lref\AALcan{E. Alvarez, L. Alvarez--Gaume and Y. Lozano, Phys. Lett.
{\bf B336} (1994) 183.}
\lref\basusy{ I. Bakas, Phys. Lett. {\bf B343} (1995) 103.}
\lref\BKO{ E. Bergshoeff, R. Kallosh and T. Ortin, Phys. Rev. {\bf D51} (1995)
3009.}
\lref\KAFK{C. Kounnas, Phys. Lett. {\bf B321} (1994) 26;
I. Antoniadis, S. Ferrara and C. Kounnas, Nucl. Phys. {\bf B421} (1994) 343.}

\lref\ALFR{ L. Alvarez--Gaume and D. Freedman, Commun. Math. Phys.
{\bf 80} (1981) 443.}
\lref\GHR{S. Gates, C. Hull and M. Rocek, Nucl. Phys. {\bf B248} (1984)
157.}
\lref\HOPAPA{P. Howe and G. Papadopoulos, Nucl. Phys. {\bf B289} (1987) 264;
Class. Quant. Grav. {\bf 5} (1988) 1647.}

\lref\zumino{B. Zumino, Phys Lett. {\bf B87} (1979) 203.}

\lref\PNBW{P. van Nieuwenhuizen
and B. de Wit, Nucl. Phys. {\bf B312} (1989) 58.}
\lref\CHS{ C. Callan, J. Harvey and A. Strominger, Nucl. Phys.
{\bf B359} (1991) 611.}

\lref\hassand{S.F. Hassan ``T--Duality and Non--local Supersymmetries'',
CERN-TH/95-98,hep-th/9504148.}
\lref\AAB{E. Alvarez, L. Alvarez-Gaume and I. Bakas, Nucl. Phys. {\bf B457} 
(1995) 3.}

\lref\IKR{I. Ivanov, B. Kim and M. Rocek, Phys. Lett. {\bf B343} (1995).}
\lref\BOVA{G. Bonneau and G. Valent, Class. Quant. Grav. {\bf 11} (1994) 1133.}
\lref\GIRU{ G. Gibbons and P. Ruback, Commun. Math. Phys. {\bf 115}
(1988) 267.}

\lref\BOFI{ C. Boyer and J. Finley, J. Math. Phys. {\bf 23} (1982) 1126;
J. Gegenberg and A. Das, Gen. Rel. Grav. {\bf 16} (1984) 817.}
\lref\HAGITO{ S. Hawking, Phys. Lett. {\bf A60} (1977) 81;
G. Gibbons and S. Hawking, Commun. Math. Phys. {\bf 66}
(1979) 291;
K. Tod and R. Ward, Proc. R. Soc. Lond. {\bf A368} (1979) 411.}

\lref\zachos{ T.L. Curtright and C.K. Zachos, Phys. Rev. Lett. {\bf 53}
 (1984) 1799;
E. Braaten, T.L. Curtright and C.K. Zachos, Nucl. Phys. {\bf B260} (1985) 630.}

\lref\zacdual{T.L. Curtright and 
C.K. Zachos, Phys. Rev. {\bf D49} (1994) 5408.}

\lref\FRTO{D.Z. Freedman and P.K. Townsend, Nucl. Phys. {\bf B177} (1981) 282.}
\lref\HS{P. S. Howe and G. Sierra, Phys. Lett. {\bf 144B} (1984) 451.}

\lref\hassanfi{S. Hassan, Nucl. Phys. {\bf B454} (1995) 86.}

\lref\spindel{Ph. Spindel, A. Sevrin, W. Troost and A.Van Proeyen, Nucl.
Phys. {\bf B308} (1988) 662; Nucl. Phys. {\bf B311} (1988/89) 465.}

\lref\WARNPN{ P. van Nieuwenhuizen and N.P. Warner, Comm. Math. Phys. {\bf 93}
(1984) 277.}

\lref\kilcom{ P. Candelas, G. Horowitz, A. Strominger and E. Witten, 
Nucl. Phys.
{\bf B258} (1985) 46; A. Strominger, Nucl. Phys. {\bf B274} (1986) 253.}

\lref\canerl{A. Giveon, E. Rabinovici and G. Veneziano, Nucl. Phys.
{\bf B322} (1989) 167;
K. Meissner and G. Veneziano, Phys. Lett. {\bf B267} (1991) 33.}

\lref\hucom{ C. Hull, Mod. Phys. Lett. {\bf A5} (1990) 1793; 
C. Hull and B. Spence, Nucl. Phys. {\bf B345} (1990) 493.}

\lref\basfewopr{ I. Bakas and K. Sfetsos, work in progress.}


\hfill {THU-95/20}
\vskip -.3 true cm
\rightline{September 1995}
\vskip -.3 true cm
\rightline {hep-th/9510034}

\bs\bs

\centerline  {\bf DUALITY AND RESTORATION OF MANIFEST SUPERSYMMETRY }

\vskip .9 true cm

\centerline  {  {\bf Konstadinos Sfetsos}{\footnote{$^*$}
{e-mail address: sfetsos@fys.ruu.nl }}                                     }

\bigskip

\centerline {Institute for Theoretical Physics }
\centerline {Utrecht University}
\centerline {Princetonplein 5, TA 3508}
\centerline{ The Netherlands }


\vskip 1.20 true cm

\centerline{ABSTRACT}

World--sheet and spacetime supersymmetries that are manifest in some string
backgrounds may not be so in their T--duals. Nevertheless, they always 
remain symmetries of the underlying conformal field theory. In previous work 
the mechanism by which T--duality destroys manifest supersymmetry and gives
rise to non--local realizations was found. We give the general conditions 
for a 2-dim $N=1$ supersymmetric $\s$-model to have non--local and hence 
non--manifest extended supersymmetry. We then examine T--duality as a 
mechanism of restoring manifest supersymmetry. This happens whenever 
appropriate 
combinations of non--local parafermions of the underlying conformal field 
theory become local due to non--trivial world--sheet effects. We present, 
in detail, an example arising from the model 
$SU(2)/U(1) \otimes SL(2,\IR)/U(1)$ 
and obtain a new exact 4-dim axionic instanton, that generalizes the 
$SU(2) \otimes U(1)$ semi--wormhole, 
and has manifest spacetime as well as $N=4$ world--sheet supersymmetry. 
In addition, general necessary conditions for abelian T--duality to preserve 
manifest $N=4$ 
world--sheet supersymmetry are derived and applied to WZW models based on 
quaternionic groups. We also prove some theorems for $\s$--models with 
non--local $N=4$ world--sheet supersymmetry.

\vskip .3 true cm

\vfill\eject


\newsec{ Introduction }

T--duality is a stringy property that provides an equivalence between strings
propagating in different backgrounds \refs{\BUSCHER,\dualrev}. Exploration of
the interplay between T--duality and other symmetries
sheds light on the role and the validity of the effective field theories 
describing them.
The most interesting such interplay is the one with supersymmetry.
It was noticed in some examples that abelian T--duality leads to 
apparent violations of extended $N=4$ world--sheet supersymmetry as well as 
of spacetime supersymmetry \basusy\ (Additional examples of that kind were 
soon provided \BKO).
However, if T--duality indeed provides an equivalence between strings in 
different backgrounds, in accordance with general ideas developed by string 
theorists over the past years, then it should not lead to a real breaking of 
other genuine symmetries such as supersymmetry. 

What has been just stated is nothing but a paradox and has a natural 
explanation. That is, in certain cases,
non--local world--sheet effects associated with the
T--duality transformation replace a local realization of supersymmetry with a 
non--local one \basfep. It is at the conformal field theory (CFT) level 
where one sees the equivalence of the various descriptions and realizations 
\basfep. This point of view was also advocated in \refs{\hassand,\AAB}.
In fact non--local realizations are something quite common and natural 
in (super)CFT. For instance,
realizations of the $N=4$ superconformal algebra using 
parafermions which are non--local objects in the sense that they have 
non--local operator product expansions (OPEs) \KAFK. 
To make the antithesis it should be pointed out that all theorems that were
proved, for instance in the context of 2--dim supersymmetric $\s$--models
\refs{\zumino\ALFR\FRTO\GHR\HS\HOPAPA\PNBW\CHS-\BOVA}, assumed  
local realizations of supersymmetry, i.e. underlying complex structures
that are local functions of the target space variables.
The reason that in some cases supersymmetry seems to be destroyed by duality
is that a non--locally realized supersymmetry cannot be distinguished from a 
lost one if anyone of these theorems is used as a criterion. 
It is one of the main purposes of
the present paper to further bridge the gap between the CFT and $\s$-model
approaches by deriving conditions for non--local realizations of 
world--sheet supersymmetry to exist and revising some of these theorems. 

There is an additional motivation from a string phenomenological point of
view. If duality breaks or, as we shall see, restores manifest
supersymmetry\foot{The term manifest 
supersymmetry is equivalent to the term locally realized 
supersymmetry in this paper.} then this
phenomenon should be incorporated in supersymmetry breaking 
scenarios considered in a string phenomenological context. To that effect it is
very interesting to uncover, and when possible relate, all 
mechanisms that lead away from or to a manifest realization of supersymmetry.

This paper is organized as follows: In section 2 we 
start from a general 2--dim $\s$-model with $N=1$ world--sheet supersymmetry 
and we find the conditions for being able to have non--locally 
realized extended supersymmetry. An exploration of the consequences is 
done in appendix A in order not to interrupt the flow of the paper.

In section 3 we develop a criterion for when it is possible to obtain a 
background with manifest $N=4$ extended supersymmetry (and target space 
supersymmetry) via a duality transformation. 
Its practical use is based on the fact that only knowledge
of one (instead of three) complex structure is required. We apply this to 
several cases with notable example being WZW models corresponding to
quaternionic
manifolds. We prove that any marginal deformation 
of them in the Cartan torus (equivalent to an $O(d,d,\IR)$ transformation) 
necessarily breaks their manifest $N=4$ world--sheet supersymmetry.

In section 4 we consider in detail
a class of models arising from duality transformation on the background for
$SU(2)/U(1) \otimes SL(2,\IR)/U(1)$. We show that when a combination of
parafermions of the CFT becomes local a manifest $N=4$ world--sheet and 
spacetime supersymmetry emerges.
This is the prototype example of what we will call duality restoration of 
manifest supersymmetry. The corresponding model is a new
axionic instanton and its spacetime interpretation is that of a generalized 
semi--wormhole. 
This model was also considered in \AAB\ in relation to what was called
dynamical restoration of manifest spacetime supersymmetry. 

We end the paper in section 5
with concluding remarks and comments on directions for feature work.

We have also written two appendices. In appendix A we prove some general
theorems for backgrounds having non--locally realized $N=4$ world--sheet 
supersymmetry and point out the differences from the cases where supersymmetry 
is realized locally. 
In appendix B we prove in a class of models that, making the moduli 
parameter dynamical (namely, coordinate dependent) is 
equivalent to performing specific duality transformations. 
This shows that, at least
in these cases, the two mechanism of restoring manifest supersymmetry are
equivalent.
 

\newsec{ General conditions for non--local realizations of supersymmetry }

The action of a 2-dim $\s$-model with $N=1$ supersymmetry is given by 
\refs{\FRTO,\HS,\PNBW}
\eqn\susye{\eqalign{
& S(x,\Psi_+,\Psi_-) = \ha \int Q^+_{\m\n} \del_+ x^\m \del_- x^\n 
+ i G_{\m\n} \Psi_+^\m \bl(\del_- \Psi_+^\n + (\Om^+)^\n_{\l\r} \del_- x^\l
\Psi_+^\r \br) \cr
&+ i G_{\m\n} \Psi_-^\m \bl(\del_+ \Psi_-^\n + (\Om^-)^\n_{\l\r} \del_+ x^\l
\Psi_-^\r \br)
+\ha R^-_{\m\n\r\l} \Psi_+^\m \Psi_+^\n \Psi_-^\r \Psi_-^\l \ ,\cr}}
where $G_{\m\n}$ and $B_{\m\n}$ are the metric and the antisymmetric tensor and
$Q^\pm_{\m\n}\equiv G_{\m\n} \pm B_{\m\n}$. 
The generalized connections are defined including the torsion 
$H_{\m\n\r}\equiv \del_{[\r} B_{\m\n]}$, 
i.e., $(\Om^\pm)^\r_{\m\n}= \G^\r_{\m\n} \pm \ha H^\r_{\m\n}$, and 
$R^\pm_{\m\n\r\l}=R^\mp_{\r\l\m\n}$ are the corresponding curvature tensors.
Notably, any background can be made $N=1$ supersymmetric. In contrast,
it is well known that extended $N=2$ supersymmetry \refs{\zumino,\ALFR,\GHR}
requires that
the background is such that an (almost) complex (hermitian) 
structure $(F^\pm)^\m{}_\n$,  
for each chiral sector, exists. The conditions to be satisfied are  
\eqn\comcond{ (F^\pm)^\m{}_\l (F^\pm)^\l{}_\n = -\d^\m{}_\n \ ,\quad
F^\pm_{\m\n} +  F^\pm_{\n\m} = 0\ ,\quad D^\pm_\m (F^\pm)^\l{}_\r = 0\ ,}
where $F^\pm_{\m\n}\equiv G_{\m\l} (F^\pm)^\l{}_\n$ and 
the generalized connections are used to define the covariant derivatives.
The first two conditions guarantee that the commutator of two new
supersymmetries gives the same translation as that of two old ones and that
no translation is generated by commuting an old and a new supersymmetry.
The third condition requires that the complex structures are covariantly 
constant and guarantee the invariance of the quadratic in the fermions terms.
Moreover, its integrability condition assures the invariance of the quartic in
the fermions terms.
Then one can check that the following invariances of the action hold
\eqn\invco{ S(x,F^+ \Psi_+, \Psi_-)=
S(x, \Psi_+ ,F^- \Psi_-) = S(x,\Psi_+,\Psi_-)\ .}
Similarly, $N=4$ extended supersymmetry \refs{\ALFR,\GHR,\PNBW}
requires that, for each sector,
there exist three complex structures  $(F_I^\pm)^\m{}_\n$, $I=1,2,3$. 
Each one of them satisfies separately \comcond,  and in addition they obey
\eqn\quaalg{ F^\pm_I F^\pm_J = -\d^{IJ} + \e^{IJK} F^\pm_K\ .}

The previous results were derived under the crucial assumption that the
complex structures are local functions of the target space variables.
This was rather unquestionable in the past, but recent work 
\refs{\basfep,\hassand} shows that non--local complex structures are equally 
acceptable in a string theoretical setting and are directly related \basfep\
to parafermions of the underlying (super)CFT corresponding to the $\s$-model 
\susye.
Therefore it is interesting to investigate the conditions under which \susye\
has non--locally realized extended supersymmetry, 
in the sense that the corresponding 
complex structures are allowed to depend non--locally
on the bosonic coordinates and on the world--sheet fermions. 
Namely, let $F^\pm=F^\pm(\vec \th,x^\m)$, where $\vec \th$ is a vectorial 
notation for $N$ functionals. A general ansatz for them, consistent with 
scaling arguments, is
\eqn\nonth{ \vec \th\equiv \int 
\bl(\vec C^+_\m \del_+ x^\m + i \vec C^+_{\m\n} \Psi_+^\m \Psi_+^\n \br)d\s^+
 \ +\
\bl( \vec C^-_\m \del_- x^\m + i \vec C^-_{\m\n} \Psi_-^\m \Psi_-^\n \br)d\s^-
 \ ,}
where the tensors $\vec C^\pm_\m$ and $\vec C^\pm_{\m\n}$ depend locally 
on the $x^\m$'s.
These, as well as the complex structures themselves, will be determined by 
requiring that the action \susye\ has still the invariances \invco.
In working out the details first we examine the vanishing of anomaly terms that
are quadratic in the fermions. This and the requirement
that the commutator of an old and a new supersymmetry does not generate a 
translation give the first two conditions in \comcond\ and 
\eqn\covno{D^\pm_\m F^\pm_{\a\b}\ +\ ( \vec C^\mp_\m \cdot \del_{\vec \th})
F^\pm_{\a\b}\ =0\ ,}
where by definition ordinary covariant derivatives do not act on the 
integrand of $\vec \th$. Its contribution has been taken into account by the 
second term in \covno.
The latter is the analog of the third condition in \comcond. We see that the 
complex structures are no longer covariantly constant in the ordinary 
sense.\foot{The non--local complex structures are covariantly constant with 
respect to 
covariant derivatives which have in their definitions ordinary derivatives 
$\del_\m$ replaced by $\del_\m + \vec C^\mp_\m \cdot \del_{\vec \th}$. 
It should also be noted that ansatz \nonth\ excludes the
possibility of path ordered (when off shell) Wilson lines.
The most general, ansatz--free, conditions on the complex structures are
\eqn\conco{
D^\pm_\m F^\pm_{\a\b} \del_\mp x^\m\ +\ \tilde\del_\mp F^\pm_{\a\b}\ =\ 0\ ,}
where the tilded world--sheet derivative acts only on the 
non--local part of the complex structure and all dependence on
world--sheet fermions is ignored. Obviously if we make the ansatz 
\nonth, conditions \covno\ are recovered.}
It remains to examine the invariance of the quartic in the fermions terms in 
\susye\ under \invco. In the case of local complex structures this term is
invariant by itself thanks to the integrability condition of the third 
equation in \comcond. 
However, in our case we get extra contributions from the variation
of the quadratic in the fermions terms in the action. The condition that the
combined result is zero can be written as 
\eqn\quartps{ R^\pm_{\m\n\a}{}^\g F^\pm_{\g\b} \ - \
R^\pm_{\m\n\b}{}^\g F^\pm_{\g\a} 
\ -\  2 \vec C^\mp_{\m\n} \cdot \del_{\vec \th} F^\pm_{\a\b} \ = \ 0\ .}
Next we consider the integrability condition of \covno\ which is 
\eqn\intepr{\eqalign{& [D^\pm_{\m}, D^\pm_{\n}] F^\pm_{\a\b} \
+\ D^\pm_{[\m} \vec C^\mp_{\n]}\cdot \del_{\vec \th} F^\pm_{\a\b} \
+ \ \vec C^\mp_{[\n}\cdot \del_{\vec \th} D^\pm_{\m]} F^\pm_{\a\b}  \cr
& = R^\pm_{\m\n\a}{}^\g F^\pm_{\g\b}\ -\ R^\pm_{\m\n\b}{}^\g F^\pm_{\g\a}\
\mp\ H_{\m\n}{}^\g D^\pm_\g F^\pm_{\a\b} \cr
& + (\del_{[\m} \vec C^\mp_{\n]} \mp H_{\m\n}{}^\g \vec C^\mp_\g)\cdot
 \del_{\vec\th} F^\pm_{\a\b}\
+\ \vec C^\mp_{[\n}\cdot \del_{\vec\th} D^\pm_{\m]} F^\pm_{\a\b}\ =\ 0\ ,\cr }}
where the second line arises from the first term of the first line and is the 
usual result one obtains from the commutator of two covariant derivatives.
The first term of the third line is an explicit rewriting of the second term 
of the first line after taking into account that only the antisymmetric part 
of the connections contributes. 
Notice that the chain rule and all the usual properties of 
the covariant derivatives are valid since, as already mentioned,
by definition they do not act on the integrand of $\vec \th$. 
Now using \covno\ to rewrite the covariant derivatives
we see that the two terms proportional to the torsion cancel each other and 
that the last term is zero. The final result is 
\eqn\integ{ R^\pm_{\m\n\a}{}^\g F^\pm_{\g\b} \ - \
R^\pm_{\m\n\b}{}^\g F^\pm_{\g\a}\  + \ (\del_{\m} \vec C^{\mp}_{\n}  - 
\del_{\n} \vec C^{\mp}_{\m} )\cdot \del_{\vec \th} F^\pm_{\a\b}\ =\ 0\ .}
A non--vanishing third term causes the complex structures not 
to commute with the generators of the holonomy group 
$(M^\pm_{\m\n})_{\a}{}^{\b} = R^\pm_{\m\n\a}{}^\b$.
Comparison of \quartps\ with \integ\ determines that
\eqn\bmndet{\vec C^\pm_{\m\n}=-{1\over 2} (\del_\m \vec C^\pm_\n - 
\del_\n \vec C^\pm_\m )\ .}
Notice that if $\vec C^\pm_\m =\del_\m \vec \L(x)$,
then $\vec C^\pm_{\m\n}=0$ and $\vec \th = \vec \L$. Thus in this case,
the complex structure is local and the supersymmetry manifest. This observation
will be the key to understand restoration mechanisms
of manifest supersymmetry, as we will soon discuss. 

In the case that  
the non--localities arise from a duality transformation with respect to a
Killing vector $\del/{\del x^0}$ one obtains, $C^\pm_\m= \pm Q^\pm_{\m0}$ 
(defined in the dual model of course) and $\th$ 
corresponds to the Killing coordinate of 
the original model that is non-locally related to the variables of its dual 
\basfep. Also the non--local complex structure should automatically satisfy
\covno\ by construction. This was proved in \hassand, where the question,
what kind of equation the non--local complex structures arising from 
the duality transformation obey, led to the analog of \covno. Similarly,
we find that for duality with respect to $N$ commuting isometries, the 
components of the vector $\vec C^\pm_\m$ are 
$(C^\pm)^a_\m = \pm Q^\pm_{\m a}$, $a=1,2,\dots , N$.

Notice that derivatives of $F^\pm$ with respect to components of $\vec \th$
are also candidates for new independent complex structures. For simplicity let
us consider the case of one functional $\th$. It is clear from \covno\
that $\del_\th F^\pm$ satisfies the same equation and some algebra shows that
so does $F^\pm\del_\th F^\pm$. It is easy to see that
$[(\del_\th F^\pm)^2 , F^\pm]=0$ and therefore after a suitable normalization
$\del_\th F^\pm $ are indeed complex structures and so are
$F^\pm \del_\th F^\pm$. This observation will be further exploited in 
appendix A.

It should be clear that from our point of view any 
mechanism that can transform $\vec \th$ into a local function would 
simultaneously restore manifest supersymmetry. One possibility would be to 
take advantage of possible isometry groups that the background might have 
and perform a T--duality transformation. Since that will transform the 
world--sheet derivatives and the fermions (we know explicitly how in the case
of abelian duality \refs{\AALcan,\basfep,\hassand} and in principle we can 
determine that for the Principal Chiral Models using \zacdual)
in the integrand of $\vec \th$ there 
is a chance, depending on the specific background, 
that the integrand becomes a total derivative on the world--sheet.
Then as it was discussed below 
\bmndet\ the complex structures in the dual theory would become local and the
supersymmetry would be manifest.
This will be demonstrated in detail with an example in section 4.
Another mechanism could be what has been called dynamical restoration of 
manifest
supersymmetry \AAB. According to the general philosophy this corresponds
to making certain moduli parameters coordinate--dependent 
\refs{\KKdyn,\Tsdyn} (for earlier work see \GSVYdyn) and then 
demanding that this drastic modification still preserves conformal invariance. 
It might happen that manifest supersymmetry is also restored in the process
(\AAB\ and subsection 4.1 of the present paper). 
Although we have no general proof, we believe that this mechanism 
is always equivalent to a restoration via particular duality transformations. 
We prove this claim for a class of one parameter moduli models, a 
particular example of which is the one considered in section 4.
For this reason we did not focus on the dynamical restoration mechanism in
this paper.


\newsec{ Conditions for manifest supersymmetry under duality }

So far we have not required that the $\s$-model action \susye\ had any special
isometries. Let us consider the case where there is one (at least) Killing
symmetry corresponding to the Killing vector $\del/{\del x^0}$ in the adapted
coordinate system, in which the background fields do not depend on $x^0$.
This need not be the case for other geometrical objects of
interest, such as complex structures \basfep. Let us assume that 
the model \susye\ has extended $N=2$ (at least) supersymmetry, which in the
adapted coordinate system is always manifest, and that 
we have determined the corresponding complex structure in each chiral sector.
We would like to derive
a necessary condition need to be satisfied for the dual model to actually 
have manifest $N=4$ extended supersymmetry. 
In practical situations
such a check will be useful since it does not require knowledge 
of any additional complex structures. 
A related problem will be to find conditions that need to be satisfied
in order that both \susye\ and its dual have manifest $N=4$
extended supersymmetry.\foot{ By means of the relation between target space
and extended world--sheet supersymmetry \kilcom\ these will also be necessary 
conditions
for having manifest target space supersymmetry.}
Let us recall that the assumption that there exist three (local) complex 
structures satisfying \quaalg\
leads to the strong conditions \PNBW
\eqn\rrff{ \tilde R^\pm_{\m\n\a\b} (\tilde F_I^\pm)^{\a\b} = 0\ ,
\quad  I=1,2,3 \ ,}
where we have written them with tildes having in mind that they should be
satisfied in the dual to \susye\ model.
We would like to express these conditions in terms of tensors defined in the
original model \susye.
The complex structures in the dual model are \refs{\IKR,\hassanfi,\basfep}
\eqn\tranc{
 (\tilde F^\pm_I)_{\m\n} = ( A_\pm^T F^\pm_I A_\pm )_{\m\n} \ , } where 
\eqn\defapm{ (A_\pm)^\m{}_\n = \bordermatrix{ & 0 & j \cr 0 & \pm G_{00}^{-1} &
- G_{00}^{-1} Q^{\pm}_{j0} \cr i & 0 & \d^i{}_j \cr }\ . }
The curvature tensors of the dual model can be extracted
from an expression in \hassand. We find
\eqn\curvte{ \tilde R^\pm_{\m\n\a\b} = (A_\mp)^\l{}_\m (A_\mp)^\r{}_\n 
(A_\pm)^\g{}_\a (A_\pm)^\d{}_\b\ \bl( R^\pm_{\l\r\g\d}\ +\ \ha G_{00}^{-1}
\del_{[\l} Q^\pm_{0\r]} \del_{[\g} Q^\mp_{0\d]} \br)\ .}
Combining \tranc\ with \curvte\ we see that \rrff\ takes the form
\eqn\rdual{ R^\pm_{\m\n\a\b} (F_I^\pm)^{\a\b}\ +\ G_{00}^{-1} 
\del_{[\m} Q^\pm_{0\n]} \del_\a Q^\mp_{0\b} (F_I^\pm)^{\a\b}\ =\ 0\ ,\quad
I=1,2,3\ .}
If the original model had manifest $N=4$ then the first term is zero and the 
condition for having manifest $N=4$ in the dual model 
reduces to the vanishing of the second term in \rdual. 
Let us emphasize that \rdual\ is only a necessary 
condition for $N=4$ supersymmetry and only its violation leads to a 
definite conclusion that the duality transformation has broken manifest $N=4$.
Nevertheless, we know of no examples where
the reverse is not also true, i.e. satisfying \rdual\ seems to always lead
to a manifestly $N=4$ supersymmetric dual model.

As examples let us briefly consider 4-dim pure
gravitational backgrounds with $N=4$ extended supersymmetry, which are known
to be hyper--kahler self--dual manifolds, 
that in addition have one Killing symmetry.
A complete classification of them exists and depends on whether or not the
covariant derivative of the corresponding Killing vector is self--dual \BOFI.
Accordingly, the Killing vector is of the translational or the 
rotational type. 
In the translational case the general forms of the 
metric and the three complex structures have been found in \HAGITO\ and 
in \GIRU\ 
respectively, whereas in the rotational case in \BOFI\ and \basfep. 
One can explicitly check that in the translational case
$\del_\a G_{0\b} F_I^{\a\b}$ is 0, for all three complex structures. 
In the rotational case, we have checked the same expression 
for the complex structure adapted to the Killing vector\foot{ A complex 
structure adapted to a Killing vector is by definition one that is a 
singlet under transformations generated by this Killing vector.}
and we found that it is 2.\foot{For the other two complex structures 
that form a doublet this expression is zero. A general explanation of this
is given in appendix A.}
Since $\del_{[\m} G_{0\n]} \neq 0$ we conclude that in general a duality 
transformation with respect to a rotational 
Killing vector cannot preserve $N=4$ as a locally realized supersymmetry. 
This is in full 
agreement with the conclusion reached in \basfep\ on the basis that 
two of the complex structures in the rotational case become non--local under 
duality. 

Next we consider marginal deformations of WZW models
based on quaternionic groups (this implies, a dimensionality that is a multiple
of four, three complex structures and manifest $N=4$; 
for a complete analysis see \spindel) 
by current--bilinears in the Cartan torus. 
This is equivalent to duality transformations \dualmargi\ 
(strictly speaking $O(d,d,\IR)$ transformations).
We will prove using \rdual\ that any such
deformation leads to a breaking of manifest $N=4$. Moreover,
with no additional effort, we will show that if a WZW model has
$N=2$ only (and a multiple of four dimensionality) 
no marginal deformation in the Cartan torus can result in a dual model 
that has manifest $N=4$.
We examine the case of one parameter duality transformations.
The proof starts by recalling that for any WZW model the generalized
curvature tensors are identically zero \zachos, i.e. $R^\pm_{\m\n\r\l}=0$,
reflecting the parallelizability of the corresponding group manifold.
Thus the second term in \rdual\ should vanish by itself for both the + and
the -- sign. One way this can be satisfied is by having $G_{00}=\const$ and
$Q^-_{0i}=0$, i.e., the duality corresponds to a chiral isometry.
However, this is not an interesting case since then the background is
self-dual \refs{\AABL,\AALcan,\hassand}. Thus for a duality with respect to any
other kind of isometry the only possibility is to 
have $\del_\a Q^\mp_{0\b} (F_I^\pm)^{\a\b} = 0$ for both signs and for 
all three complex structures. 
The simplest type of isometry corresponding to a non-trivial duality
transformation is a mixed one of chiral and anti-chiral type. Specifically we
parametrize the group element $g\in G$ as 
\eqn\paragroup{g=e^{i \th_L T} h(x) e^{i \th_R T}\ ,\quad
\th_L=(Q+1) \tau + \psi\ ,\quad \th_R=(1-Q)\tau -\psi\ ,}
where $T$ is a Cartan generator and $Q$ is a constant modulus. 
Then the corresponding WZW action \Wwzw\ $S=k I_{wzw}(g)$ takes the form
\eqn\debo{\eqalign{ I_{wzw}(g) = 
& I_{wzw}(h)\ +\ {1\ov \pi} \int \bl((1+Q^2) + (1-Q^2)\S
\br)\del_+\tau \del_-\tau \ + \ (1-\S) \del_+\psi \del_-\psi \cr
&\ +\ \bl((Q-1) - (1+Q) \S \br) \del_+\tau \del_-\psi  
\ +\ \bl((1+Q) + (1-Q) \S \br) \del_+\psi \del_-\tau \cr
&\ +\ \bl( (1+Q) \del_+ \tau +\del_+ \psi \br) J^-_i \del_- x^i \ 
-\ J^+_i \del_+ x^i \bl( (Q-1) \del_- \tau +\del_- \psi \br) \ ,\cr }}
where we used the definitions ($\del_i \equiv {\del/\del x^i}$)
\eqn\jjs{ J^+_i= -i \Tr(T h\inv \del_i h)\ ,\quad 
J^-_i= -i \Tr(T \del_i h h\inv)\ ,\quad \S = \Tr(T h T h\inv )\ .}
The action \debo\ will be dualized
with respect to the Killing vector $\del/{\del \tau}$ and the
matrix $Q^\pm_{\m\n}$ can be easily read off. 
Notice that, if the modulus $Q=\pm 1$ then the isometry becomes chiral. 
Assuming that $Q\neq \pm 1$ and using the complex structures $F_3^\pm$
adapted to the Killing vector $\del/{\del\tau}$,
the conditions $\del_\a Q^\mp_{0\b} (F_3^\pm)^{\a\b} = 0$ become
\eqn\dqff{  \del_i \S \bl( (Q\pm 1)  (F_3^\pm)^{\tau i}  +  
(F_3^\pm)^{\psi i} \br)\  \mp \  \del_i J^\pm_j (F_3^\pm)^{ij}  =0\ .}
The above equations cannot be satisfied, as we shall show next.
Let $T^A=\{T^1,T^a\}$, $a=2,3,\dots , \dim(G)$ denote the Lie algebra 
generators with $T^1$ the generator in the Cartan subalgebra we have been 
using, i.e., $T^1\equiv T$. It is also convenient to define
\eqn\ddll{ L^A_\m= -i \Tr(T^A g\inv \del_\m g)\ ,\quad 
R^A_\m= -i \Tr(T^A \del_\m g g\inv)\ ,\quad C^{AB} = \Tr(T^A g T^B g\inv )\ ,}
with \jjs\ related to them in an obvious way.
Then a straightforward computation gives
\eqn\dernan{\eqalign{ & \del_i J^+_j (F_3^+)^{ij} = 
\bigl( \ha (f^+)^{a b} - 
 L^a_\tau L^b_i (F_3^+)^{\tau i} -  L^a_\psi L^b_i (F_3^+)^{\psi i}
- L^a_\tau L^b_\psi (F_3^+)^{\tau\psi} \br) f^1_{a b}\ ,\cr
& \del_i \S =- C^{1 a} L^b_i f^1_{a b} \ ,\quad L^a_\tau =(1+Q) C^{1 a}\ ,
\quad L^a_\psi = C^{1 a} \ , \cr }}
where $f^1_{a b}$ is the relevant Lie algebra structure constant.
Note that, $(f^+)^{a b}$ is a Lie algebra complex structure \spindel,
defined as $(F_3^+)_{\m\n}= L^A_\m L^B_\n (f^+)_{A B}$.
Using \dernan\ and its analog in the other chiral sector, the conditions 
\dqff\ take the simplified form $f^1_{a b} (f^\pm)^{a b}=0$.
In the Cartan basis the indices $a,b$ run only over pairs of
positive and negative roots ($\a,\bar\a$) of the Lie algebra
(if either $a$ or $b$ takes a value 
in the Cartan subalgebra then the corresponding structure constant vanishes).
In this basis $(f^\pm)^\a{}_\a=- (f^\pm)^{\bar\a}{}_{\bar\a}= i$ \spindel\ 
and the conditions become
$\sum_\a f_{\a 1}{}^\a  \sim \sum_\a \a^1 = 0$, where $\a^1$ denotes the 
component of the positive root in the direction of the Killing vector.
Clearly this equality cannot be satisfied.

Thus, we have proved that a one parameter marginal deformation in the Cartan
torus of a WZW model for a quaternionic group breaks its
manifest $N=4$ extended supersymmetry. 
If the original WZW model had only an $N=2$ we also conclude that 
duality cannot enlarge the supersymmetry to an $N=4$. Obviously,
the same conclusions are valid for more complicated $O(d,d,\IR)$
transformations. 

Interesting situations arise when none of the terms in \rdual\ is zero but
their forms are such that the equality is satisfied. A non--vanishing 
first term implies that only the $N=2$ supersymmetry was manifest in the 
original model and that there were two additional ones, non--locally 
realized. 
It is only after duality we are actually promoting it to a manifest
$N=4$. This is what we call duality restoration of manifest supersymmetry.

\newsec{ An explicit example }

In order to demonstrate every aspect related to the 
general discussion of the previous sections we consider 
the following background with line element given by
\eqn\met{ ds^2 = d\vphi^2 + \cot^2{\vphi}\ dx^2 + d\r^2 + R^2(\r)\ dy^2\ ,}
and zero antisymmetric tensor. It is well known and easy to compute that 
1--loop conformal invariance requires that a non--trivial dilaton is induced
with
\eqn\dil{ \Phi= \ln(\sin^2\vphi/R'(\r))\ , }
where the function $R(\r)$ satisfies the differential equation
\eqn\condR{ R'=C_1 R^2 + C_2 \ ,}
whose general solution is easy to obtain (see for instance \KKdyn).
Depending on the constants $C_1,C_2$ the solutions
correspond to a 4--dim model that can be considered as the tensor product of 
the coset CFT model
$SU(2)/U(1)$ with another $SU(2)/U(1)$ model (if $C_1C_2>0$) or with 
$SL(2,\IR)/U(1)$ (if $C_1C_2<0$) or with 2--dim
flat space (if $C_1=0$) or with the dual to 2--dim flat space (if $C_2=0$).
The central charge deficit from the classical value $c=4$ is $\d c\sim -\a'(1+
C_1C_2)$.
Next we introduce the coordinate change
\eqn\coord{ x= \psi -{\tau\ov 2}\ ,\quad y=\psi +{\tau\ov 2} }
and perform a T--duality transformation with respect to the symmetry
generated by the $\del/{\del\psi}$ Killing vector, thus obtaining the
background
\eqn\dubac{\eqalign {
&d\tilde s^2 = d\r^2 + d\vphi^2 +{1\ov 1 + R^2 \tan^2\vphi }
( \tan^2\vphi\ d\tilde\psi^2 + R^2 \ d\tau^2 )\ ,\cr
&\tilde B_{\tau\tilde\psi}= {1\ov 1+ R^2 \tan^2\vphi}\ ,\quad
\tilde \Phi = \ln\bl((\cos^2\vphi + R^2 \sin^2\vphi)/R' \br)\ ,\cr } }
where we have denoted the dual to $\psi$ variable by $\tilde\psi$.
Their explicit relation follows in the formulation of abelian T--duality as a
canonical transformation \AALcan\ (see also \refs{\canerl,\zacdual}) and 
can be found using general formulae that relate world--sheet derivatives 
\basfep
\eqn\reltt{ \del_\pm \psi={1\ov 1+R^2 \tan^2\vphi}
\bl(\pm \tan^2\vphi\ \del_\pm\tilde\psi 
+\ha ((1-R^2 \tan^2\vphi)\ \del_\pm \tau) \br)\ .}
This formula will be very important as we shall shortly see.

We now turn to the question of world--sheet supersymmetry for the 
background
\met. There is only one local complex structure that 
solves \comcond\ given by
\eqn\colo{F_3= \cot \vphi\ d\vphi \wedge dx + R(\r)\ d\r \wedge dy \ ,}
for any function $R(\r)$. Notice that, there is no distinction between the +
and the -- components since the antisymmetric tensor is zero.
However if the locality condition for the complex structures is relaxed one
can search for solutions of \covno\ and one finds that there are two for
each chiral sector. In order to present them it is first
convenient to introduce the parafermionic type, 1--forms
\eqn\formse{\eqalign{& \Psi^{(1)}_\pm = (d\vphi\ \pm \ i \cot\vphi\ d x)\
e^{\pm i (-x + \th_1)}\ , \quad \bar \Psi^{(1)}_{\pm}  = 
(d \vphi\ \mp \ i \cot\vphi\ d x)\ e^{\pm i (x + \th_1)}\ ,\cr
& \th_1\equiv \int \cot^2\vphi \del_+ x d\s^+ - \cot^2 \vphi \del_- x d\s^- \ ,
\cr } }
and 
\eqn\formee{\eqalign{& \Psi^{(2)}_\pm = (d \r\ \pm\ i R\ d y)\
e^{\pm i (c_2 y + \th_2)}\ ,\quad \bar \Psi^{(2)}_{\pm}  = 
(d \r\ \mp\ i R\ d y)\ e^{\pm i (-c_2 y + \th_2)} \ ,\cr 
&\th_2\equiv  \int (c_2- R') \del_+y  d\s^+ - (c_2-R')\del_-y d\s^-\ ,
\cr } }
where $c_2$ is an arbitrary constant.
These have a natural decomposition in terms of $(1,0)$ and $(0,1)$ forms on the
string world--sheet
\eqn\decof{ \Psi^{(a)}_\pm =\Psi^{(a)}_{\pm,+} d\s^+ \ + \ 
\Psi^{(a)}_{\pm,-} d\s^- \ ,\quad 
\bar\Psi^{(a)}_\pm = \bar\Psi^{(a)}_{\pm,+} d\s^+ \ + \
\bar\Psi^{(a)}_{\pm,-} d\s^- \ ,} 
where $a=1,2$. It can be easily verified using the classical equations of
motion for the model \met\ that the chiral and anti--chiral conservation laws 
\eqn\onss{ \del_- \Psi^{(a)}_{\pm,+}=0\ ,\quad 
\del_+ \bar\Psi^{(a)}_{\pm,-}=0 \ ,}
are obeyed. In fact in this case $\Psi^{(1)}_{\pm,+}$ and 
$\bar \Psi^{(1)}_{\pm,-}$ 
are nothing but the classical parafermions for the $SU(2)/U(1)$ coset. 
However, $\Psi^{(2)}_{\pm,+}$ and 
$\bar\Psi^{(2)}_{\pm,-}$,
even though are of the parafermionic type, cannot be identified with 
classical parafermions of some 2--dim CFT unless
the function $R(\r)$ obeys \condR, i.e., the corresponding $\s$--model is 
conformal.
Using the above definitions the expressions for the two  non--local 
complex structures are 
\eqn\nolocs{ F_1^+ = \Psi^{(1)}_+ \wedge \Psi^{(2)}_+ \ +\ \Psi^{(1)}_- \wedge
\Psi^{(2)}_- \ ,\qq
F_2^+ = i \Psi^{(1)}_+ \wedge \Psi^{(2)}_+\ -\ i  \Psi^{(1)}_- \wedge
\Psi^{(2)}_- \ ,}
and 
\eqn\noloscc{ F_1^- = \bar\Psi^{(1)}_+ \wedge \bar\Psi^{(2)}_+ \
+\ \bar\Psi^{(1)}_- \wedge \bar\Psi^{(2)}_- \ ,\qq
F_2^- = - i \bar\Psi^{(1)}_+ \wedge \bar\Psi^{(2)}_+ \ + \
i \bar\Psi^{(1)}_- \wedge \bar\Psi^{(2)}_- \ ,}
where we have neglected writing explicitly the necessary for non--local
complex structures dependence on the world--sheet fermions since 
it is completely fixed by the bosonic part (see \nonth\bmndet).
Notice that, there is a distinction between the + and the -- components even 
though the antisymmetric tensor is zero. This is a new feature that can only
happen in non--local realizations of extended supersymmetry.
It is interesting that the local complex structure $F_3$,
can also be written in terms of the parafermionic 1--forms as
\eqn\ftloc{ F_3 = i \Psi^{(1)}_+ \wedge \Psi^{(1)}_- 
\ +\  i \Psi^{(2)}_+ \wedge \Psi^{(2)}_- \ = \
- i \bar\Psi^{(1)}_+ \wedge \bar\Psi^{(1)}_- \
-\ i \bar\Psi^{(2)}_+ \wedge \bar\Psi^{(2)}_-  \ ,}
where a simple inspection shows that both alternative expressions reduce 
to that in \colo. The structure of the non--locally realized $N=4$ we have
just exhibited in this example is in full agreement with general conclusions
in appendix A. 

We now turn to the question of world--sheet supersymmetry for the dual
background \dubac.
Instead of solving the corresponding conditions \covno\ we apply the general
formula \tranc\ in our case (first we pass to the coordinate system \coord)
and in addition for the non-local complex structures $F^\pm_1$ and $F^\pm_2$
we use the transformation rules for the world--sheet derivatives \reltt\
in order to deduce the transformation of the functionals $\th_1,\th_2$
defined in \formse\formee.
The dual of the local complex structure $F_3$, in each chiral sector, is 
\eqn\colod{\tilde F^\pm_3 = {1\ov 1+ R^2 \tan^2\vphi}\ \bl(R d\r \wedge
(d\tau\ \pm\ \tan^2\vphi d\tilde\psi) \ +\  \tan\vphi (R^2 d\tau \  
 \mp\  d\tilde\psi) \wedge d \vphi )\ .}
Indeed it can be explicitly verified that this is a complex structure for the
dual model for any function $R(\r)$. Next we consider the transformation
of the non--local complex structures \nolocs\noloscc. 
From \formse\formee\ we see that generically the non--localities 
will persist in the corresponding complex structures of the dual model which
will also have non--locally realized $N=4$ supersymmetry.
However, a closer 
look reveals that there is a particular case where they completely cancel out 
after the duality is performed.  Let us consider the  phase factors 
in $F^\pm_{1,2}$ where all non--localities lie,
\eqn\phas{\eqalign{ &\th_1 + \th_2 \ \pm\ (c_2 y -x) \ = \
\pm (c_2 -1) \psi \ \pm\  \ha (c_2+1) \tau \cr
& + \int \bl( (c_2-R' + \cot^2\vphi)\del_+ \psi \ + \ \ha 
(c_2-R' - \cot^2\vphi)\del_+\tau \br) d\s^+\ -\ \bl(\ +\to -\ \br)\ ,\cr} }
where we have passed to the coordinate system \coord.
Under duality with respect to $\psi$ the world--sheet derivatives
$\del_\pm\psi$ will transform as in \reltt. It is easily now seen that only
if we choose $c_2=1$ and the function $R(\r)$ to satisfy
\eqn\eqnno{R'=1-R^2 \quad \Rightarrow \quad R=\tanh\r \ \ {\rm or}\ \
 \coth \r \ ,}
all non--localities in the dual complex structures disappear.\foot{ Equation
\eqnno\ also arises by demanding that \rdual\ is satisfied for the background
\met\ and the complex structure \colo\ (in the coordinate basis \coord\
with $x^0\equiv \psi$). Also we have omitted the obvious solution $R=1$ that
corresponds to the $SU(2)\otimes U(1)$ WZW model. Nevertheless, all of our 
formulae below that contain $R$ explicitly will be valid for $R=1$ as well.
What is 
important to mention is that a marginal deformation away from the WZW point
($R=\const \neq 1$) leads to a loss of manifest $N=4$, 
in agreement with the general statement of section 4 for WZW models based on 
quaternionic groups.}
Indeed then the
phase factors \phas\ transform under duality to just $\tilde \psi \pm \tau$.
The corresponding local complex structures dual to \nolocs\noloscc\ are then
\eqn\comdd{ \pmatrix{\tilde F^\pm_1 \cr \tilde F^\pm_2}=
\pmatrix{ \cos(\tau \pm \tilde\psi)
& \sin(\tau \pm \tilde \psi) \cr -\sin(\tau\pm \tilde\psi)
& \cos(\tau \pm \tilde \psi)\cr}\pmatrix{f^\pm_1 \cr f^\pm_2} \ ,}
with the definitions 
\eqn\tform{\eqalign{ & f^\pm_1= - d\r\wedge d\vphi\ \pm \
{R\tan\vphi \ov 1+ R^2 \tan^2\vphi}\ d\tau \wedge d\tilde\psi \cr
& f^\pm_2= {1\ov 1+ R^2 \tan^2\vphi}\ \bl(- \tan\vphi\  d\r \wedge (R^2 d\tau
\ \mp \ d\tilde \psi) \ + \ R (d\tau\ \pm\ \tan^2\vphi\ d\tilde \psi)\wedge 
d\vphi \br) \ ,\cr } }
where the function $R$ assumes either one of the two expressions in
\eqnno. Notice that, in agreement with what was expected for rotational--type
Killing vectors \refs{\GIRU,\basfep} (see also appendix A), out of the 
three complex 
structures, $\tilde F^\pm_3$ is a singlet of the duality group 
($SO(2)$ in this case), whereas $\tilde F^\pm_1$ and $\tilde F^\pm_2$ form a 
doublet, in each chiral sector separately.

It is important to emphasize that in trying to obtain a 2-dim $\s$-model with
manifest $N=4$ supersymmetry from \met\ via a duality transformation
at no point we required that \met\ or its dual was conformally invariant. The 
entire treatment was completely classical and 
the function $R(\r)$ remained arbitrary. 
Both \met\ and its dual \dubac\ have non--locally realized $N=4$ 
supersymmetry at the classical level.
It turned out that the condition \eqnno\ that led to manifest $N=4$
supersymmetry for the dual model is a particular case of \condR, with
$C_2=-C_1=1$, that guarantee 1--loop conformal invariance for both models. 
For these choices for $R(\r)$, \met\dil\ correspond to the direct product
$SU(2)/U(1)_k \otimes SL(2,\IR)_{-k-4}/U(1)$ and 
the central charge deficit is zero. Also $\Psi^{(2)}_{\pm,+}$ and 
$\bar \Psi^{(2)}_{\pm,-}$ become the usual 
classical non--compact parafermions of the $SL(2,\IR)/U(1)$ coset.

\subsec { Spacetime Supersymmetry }

In \AAB\ a partial proof was given that the model \dubac\ with the function
$R(\r)$ satisfying \eqnno\ has manifest spacetime supersymmetry by showing
that only with these choices the dilatino equation can be satisfied. 
In view of possible 
subtleties \WARNPN\ we complete the proof of \AAB\ by solving the gravitino 
equation and finding the corresponding Killing spinors. 

The Killing spinor equations are
\eqn\kilspi{\eqalign{
& \d \Psi_\m = \bl( \del_\m + {1\ov 4} (\om_\m{}^{\a\b} -\ha H_\m{}^{\a\b} )
 \g_{\a\b} \br) \xi =0\ ,\cr
& \d \l = - \bl(\g^\m \del_\m \Phi + {1\ov 6} H_{\m\n\l} \g^{\m\n\l} \br)\xi =0
\ ,\cr }}
where $\Psi_\m$ and $\l$ are the gravitino and dilatino fields respectively
and the dilaton $\Phi$ is given by \dil.
We find that for the background \dubac\ with the choice for $R=\tanh \r$ or 
$R=\coth \r$ the Killing spinor is
\eqn\solspi{ \pmatrix{\xi_+ \cr \xi_-} = e^{-i(a_1 \s_1 + a_3 \s_3)(\tau + 
\psi) } e^{- i a_2 \s_2} \pmatrix{0 \cr \e_- }\ ,}
where $\e_-$ is the non-zero Weyl component of a constant spinor and 
\eqn\defspi{\eqalign{ & a_1 = \ha R\tan\vphi (1+ R^2 \tan^2\vphi)^{-1/2}\ ,
\quad a_3= \ha (1+ R^2 \tan^2\vphi)^{-1/2} \ ,\cr
& a_2= \ha \tan^{-1} (R \tan\vphi)\ .\cr }}
Notice that contrary to the case of restoration of manifest world--sheet 
supersymmetry which required no quantum input at all (conformal
invariance was not even an issue), restoring manifest target space
supersymmetry demanded the use of the dilaton $\Phi$ which is a 1--loop
quantum effect in the $\a'$--expansion.


\subsec{ Interpretation as a gauged WZW model }

It is useful to associate the background \dubac\ for the special cases where 
$N=4$ extended world--sheet and spacetime supersymmetry become
manifest, with a gauged WZW model. Consider the gauged WZW type action
\eqn\gwzw{ S= k I_{wzw}(h_+\inv g_1 h_-) - k I_{wzw}(h_+\inv g_2 h_-)\ , }
for the group elements $g_1\in SU(2)$ and $g_2\in SL(2,\IR)$ parametrized as
\eqn\paramm{ g_1 = e^{i \s_1 \th_L } e^{i \s_3 \vphi} e^{i \s_1 \th_R}
\ ,\qq g_2 = e^{i \s_1 \om_L } e^{ \s_3 \r} e^{i \s_1 \om_R} \ ,}
and where $h_\pm$ are two $U(1)$ group elements that parametrize the gauge 
fields $A_\pm \equiv h_\pm\inv \del_\pm h_\pm $.
Notice that the action \gwzw\ does not contain the typical for gauged
WZW models term $I_{wzw}(h\inv_+ h_-)$. 
Nevertheless, as we shall see bellow, the gauge field dependence of \gwzw\ 
is expressible in terms of gauge fields in a local way even without 
such a term.
The action \gwzw\ is manifestly invariant under the 2--parameter finite 
gauge transformation 
\eqn\gautra{\eqalign{&\d\th_L 
= \d\om_L = \e_L\ ,\quad \d \vphi=0\ , \qq h_+ \to  e^{i \s_1 \e_L} h_+ \cr
& \d\th_R= \d \om_R = \e_R \ ,\quad \d \r=0\ ,
\qq h_- \to e^{- i \s_1 \e_R} h_- \ ,\cr }}
with gauge parameters $\e_{L,R}=\e_{L,R}(\s^+,\s^-)$.
The gauge choice $\om_L=\om_R=0$ completely fixes the gauge. 
Using the Polyakov--Wiegman formula and changing variables as
$\th_L=\ha (\tau -\psi)$ and $\th_R=\ha (\tau +\psi)$ the action \gwzw\ 
takes the form
\eqn\gwzwop{\eqalign{  S= & {k\ov \pi} \int \del_+ \vphi \del_-\vphi +
\del_+ \r \del_-\r + \cos^2\vphi \del_+\tau \del_-\tau 
+ \sin^2\vphi \del_+\psi \del_-\psi \cr
& + \ha \cos 2\vphi (\del_+\tau \del_- \psi - 
\del_+\psi \del_- \tau )  +  2 i A_+ ( \cos^2\vphi \del_- \tau - \sin^2\vphi
 \del_-\psi) \cr
& - 2i ( \cos^2\vphi \del_+ \tau + \sin^2\vphi \del_+\psi)  A_- 
+ 2 A_+ A_- (\cos 2\vphi - \cosh 2\r )\ .\cr }}
It is a 
standard straightforward procedure to integrate out the gauge fields
and obtain a $\s$-model action (the non--trivial dilaton is also induced). It
turns out that this model is equivalent to \dubac, with $R=\coth \r$.
Notice that in \gwzwop\ as a consequence of \gwzw\ there is no 
$A_+ A_-$--term with constant coefficient. 
This is a characteristic of what is known as
``chiral'' gauged WZW models \refs{\tye,\KSTh}. 
We believe that the type of gauging \gwzw\
may lead to other models with manifest $N=4$ supersymmetry.


\subsec{ The spacetime }

In order to obtain a clear geometrical picture it is convenient to use 
Cartesian coordinates 
\eqn\cooch{\eqalign{ &x_1=r_0 \sinh\r \cos\vphi \cos\tau\ ,\quad 
x_2=r_0 \sinh\r \cos\vphi \sin\tau \ ,\cr
& x_3=r_0 \cosh\r \sin\vphi \cos\tilde\psi ,\quad 
x_4= r_0 \cosh\r \sin\vphi \sin\tilde\psi \ ,\cr}}
if $R=\tanh \r$ and similarly if $R=\coth \r$, where $r_0$ is an 
arbitrary radial parameter. Then the background \dubac\ takes the form 
\eqn\wormd{\eqalign{& ds^2 = e^{-\Phi}\ dx_i dx_i\ ,
\quad  H_{ijk} = -  \e_{ijk}{}^l \del_l \Phi\ ,\cr
& \Phi =  \ha\ln\bl((x_ix_i+r_0^2)^2 - 4 r_0^2 (x_3^2+x_4^2)\br)\ ,\cr }}
%
where we have omitted the tildes and for convenience we have presented the 
expression for the antisymmetric field strength, instead of the tensor itself.
The metric is conformally flat with the conformal
factor satisfying the Laplace equation adapted to the flat metric, 
i.e. $\del_i\del_i e^{-\Phi}=0$,
in agreement with a general theorem proved in \CHS.\foot{This theorem states
that in 4--dim backgrounds with $N=4$ extended world--sheet supersymmetry 
and torsion, the metric is conformally related 
to a hyper--kahler one with the conformal factor satisfying the Laplace
equation as defined using the hyper--kahler metric. This 
conclusion is not always true in cases with $(4,0)$ world--sheet supersymmetry 
\BOVA. Nevertheless, it is true in our case where we have $(4,4)$ world--sheet
supersymmetry.}
The antisymmetric field strength solves the (anti)self--duality conditions 
of the dilaton--axion field and therefore our solution \wormd\ is an 
axionic--instanton.
For completeness we write down the form of the complex structures in the 
coordinate system \cooch
\eqn\conew{\eqalign{
& F^\pm_1 = e^{-\Phi}(- dx_1 \wedge dx_3 \pm dx_2\wedge dx_4)\ ,\cr
& F^\pm_2 = e^{-\Phi}(\pm dx_1 \wedge dx_4 + dx_2\wedge dx_3)\ ,\cr
&F^\pm_3 = e^{-\Phi}( dx_1 \wedge dx_2 \pm dx_3\wedge dx_4) \ .\cr }}
In fact these are complex structures for all 4--dim axionic instantons of the 
form \wormd\ irrespectively of the particular dilaton $\Phi$.

Geometrically the metric in \wormd\ represents a semi--wormhole with a fat
throat.\foot{Actually it represents the throat of the wormhole itself. 
A true semi--wormhole is obtained only by shifting $e^{-\Phi}$ by a 
constant since then asymptotically the space is Euclidean. In our case this 
corresponds to an $S$--duality transformation. 
The same remarks hold for the $SU(2)\otimes
U(1)$ semi--wormhole (see below).}
The metric has singularities not at a single point, 
but in the ring $x_1=x_2=0$, $x_3^2+x_4^2=r_0^2$
(obviously the radius of the ring can be set equal to 1 by an overall
rescaling of the coordinates). Therefore the throat never becomes infinitely
thin. 
In the region around the origin at $x_i=0$ (equivalently, if we let the ring
radius become very large $r_0\to \infty$)
the background \wormd\ becomes that corresponding to flat space with constant
dilaton and antisymmetric tensor. Far away from the ring we expect that its
presence should not play any role.
Indeed, if we let $x_i=y_i/\e$ and $\e\to 0$ (equivalently if we let 
$r_0\to 0$) we obtain a solution of the 
form \wormd\ but with $\Phi= \ln(y_i y_i)$. 
%
%
This is exactly the background of the usual semi--wormhole corresponding to 
the $SU(2)\otimes U(1)$ WZW model (with a background charge) \refs{\CHS,\KAFK}.
Note that due to the ring singularity structure the isometry group
of the usual semi--wormhole metric $O(4)$ breaks down to 
$SO(2)\otimes SO(2)$ for our semi--wormhole.
Let us finally mention that close to the singularity ring the isometry group of
the metric is enhanced to $SO(3)$. This can be seen by letting $x_i=2\e y_i$,
for $i=1,2,4$, $x_3=r_0 + 2 \e y_3$ and then taking the limit $\e\to 0$ 
and absorbing a factor $\e$ into a redefinition of the string coupling $\a'$
(at the CFT level this corresponds to a contraction).
The resulting axionic instanton is again of the form \wormd\ but with 
$\Phi=\ha \ln(y_1^2+ y_2^2 + y_3^2 )$, thus revealing the advertised 
$SO(3)$ isometry. It can be shown that this space is duality related to flat
space.

The background \wormd\ is the most general axionic instanton in 4--dim for 
which there is 
manifest $N=4$ supersymmetry and the corresponding CFT is known. 
It encompasses every other similar 
solution that has appeared in the literature so far \KAFK\ since they are
either duality related to it or they can be obtained from it via a combination 
of duality and 
contraction procedures of the type we have described above. 


\newsec{ Concluding remarks }

In this paper we provided general conditions for the existence of non--locally
realized extended world--sheet supersymmetry in 2--dim supersymmetric 
$\s$-models. This has implications for the background fields.
For instance, we prove in appendix A, among other theorems,
that $N=4$ (realized non--locally) does not imply that the manifold is Ricci 
flat in the absence of torsion. Next we examined the 
question of restoring manifest (equivalently locally realized) supersymmetry
via duality transformations and gave general necessary conditions for being
able to do that. Such restoration happens when due to non--local world--sheet 
effects taking place in the duality transformation,
 the non--local complex structures become local. This is the reverse
mechanism of that destroying manifest supersymmetry. 
In the case that the underlying
CFT is known the non-localities are better represented by the usual 
parafermionic objects. This was explicitly demonstrated in a new 4--dim axionic
instanton representing a semi--wormhole (generalizing the one
corresponding to $SU(2)\otimes U(1)$) with a fat throat. 
The manifest spacetime supersymmetry of this background was also 
explicitly demonstrated by solving the Killing spinor equations.

The emergence of a manifest $N=4$ raises 
some interesting questions concerning realizations of the
$N=4$ superconformal algebra.
It even suggests the existence of new type of parafermions that 
are the fundamental symmetry generating objects in these backgrounds. 
Specifically, since our axionic instanton came as a particular duality
transformation on the background corresponding to the tensor product 
$SU(2)_k/U(1) \otimes SL(2,\IR)_{-k-4}/U(1)$, the starting point in any
realization of the $N=4$ superconformal algebra would be to use the 
corresponding compact and non--compact parafermions \KAFK.
However, as we have seen in each chiral sector one combination 
of these parafermions becomes a local object as a manifest $N=4$ emerges. That
suggests that only a combination of them, orthogonal to the first,
behaves in a non--local manner, i.e., is a true parafermion. 
Since this was demonstrated only classically one should try to 
further develop this idea at the CFT level.

We believe that the general conditions for existence of non--local extended 
world--sheet supersymmetry we have presented in this paper
should be used to explicitly
demonstrate the hidden non--local supersymmetries of models obtained via 
non--abelian duality transformations \nadual. 
Prototype examples are 4--dim $SO(3)$--invariant hyper--kahler metrics.
In these cases the
non-abelian duality is performed with respect to the left (or right action)
of the isometry group $SO(3)$. For a class of such metrics,
that includes the Taub-NUT and the Atiyah-Hitchin,
the three complex structures 
transform in the triplet representation of $SO(3)$ \GIRU.
Non--abelian duality will break the original $N=4$ as a local symmetry 
completely down to an $N=1$ and a non--local realization 
will emerge \basfep. 
It turns out that path ordered exponentials appear 
and the relevant equation to consider then is \conco.
Work along these lines is in progress.

It is also conceivable that duality and possibly dynamical restoration
of manifest supersymmetry techniques would provide a natural explanation of a 
phenomenon observed in \hucom\ in $(2,0)$ 
supersymmetric $\s$--models with one Killing symmetry and a complex structure
non--preserved under diffeomorphisms.
There, in order to close the symmetry algebra
one had to ``postulate'' a compensating 
transformation for the complex structure. We are also convinced that 
restoration of manifest supersymmetry of the type we have exhibited in this
paper can happen in more general Kazama-Suzuki models.
We hope to report along these lines in the future \basfewopr.

We believe that the ideas and techniques developed in this paper could be used 
to explore the possibility that various solutions that are of interest in
black hole physics or cosmology might have hidden supersymmetries or be 
related to solutions with manifest supersymmetry.
Since this necessarily involves non--local world--sheet effects
it would be important in our effort to understand the way string theory 
could resolve fundamental problems in physics.

\bs\bs 

\centerline { \bf Acknowledgments }

I would like to thank I. Bakas, B. de Wit and F. Hassan for many useful 
discussions during the course of this work. 
I am also grateful to the Theory Division at
CERN, where part of this work was done, for financial support and hospitality
and to the Physics Department of the University of Patras for 
hospitality during the final stage of typing this paper.


\appendix A { Theorems for non-locally realized $N=4$ }

In this appendix we prove some basic theorems for backgrounds having
non--local $N=4$ extended supersymmetry of a type that can be
described by the ansatz \nonth.

Suppose that we have three complex structures 
(non--local in general) obeying \quaalg. 
Then each one of them satisfies \covno\ and its integrability condition
\integ\ which we rewrite in a slightly different form after we multiply 
with a complex structure 
\eqn\quatrss{ R^\pm_{\m\n\a\b} = R^\pm_{\m\n\g\d} (F_I^\pm)^\g{}_\a 
(F_I^\pm)^\d{}_\b \ - \ 2 \vec C^{\mp}_{\m\n} \cdot (\del_{\vec \th} F_I^\pm
F_I^\pm)_{\a\b}\ ,}
where $I$ is being kept fixed and 
$\vec C^{\pm }_{\m\n}= -\ha \del_{[\m} \vec C^{\pm }_{\n]}$ as in \bmndet.
Then we contract \quatrss\ by $(F^\pm_J)^{\a\b}$ to obtain 
\eqn\thee{ R^\pm_{\m\n\a\b} (F_J^\pm)^{\a\b}= - R^\pm_{\m\n\a\b} (F_I^\pm
F_J^\pm F_I^\pm)^{\a\b}\ + \ 2 \vec C^{\mp }_{\m\n} 
\cdot ( \del_{\vec \th} F_I^\pm F_I^\pm F_J^\pm )_{\a}{}^{\a}\ ,}
with fixed $I,J$. After using \quaalg\ we finally obtain 
\eqn\thet{ R^\pm_{\m\n\a\b} (F_J^\pm)^{\a\b} =
R^\pm_{\m\n\a\b} (F_I^\pm)^{\a\b} \d^{IJ}\ +\
 \vec C^{\mp }_{\m\n} 
\cdot (\del_{\vec \th} F_I^\pm F_K^\pm)_{\a}{}^\a \e^{IJK}\ ,}
where we sum only over $K$.
If $I= J$ this formula is trivial. Taking $I\neq J$ we obtain
\eqn\thett{ R^\pm_{\m\n\a\b} (F_J^\pm)^{\a\b} =
\vec C^{\mp }_{\m\n} \cdot  
(\del_{\vec \th} F_I^\pm F_K^\pm)_{\a}{}^\a \e^{IJK}\ ,\quad I\neq J\ .}
In the case of manifest $N=4$ the right hand side is zero and we obtain \rrff.

Let us consider the torsionless case where the two generalized
curvatures reduce to the Riemannian one. It is well established that
then $N=4$ supersymmetry implies Ricci flatness (see for instance \PNBW). 
We will see however that this is not the case for non--local $N=4$. 
Using the cyclic identity 
$R_{\m[\n\a\b]} = 0$ we obtain 
\eqn\voith{R_{\m\n\a\b} (F_I^\pm)^{\n\b}= \ha R_{\m\a\b\n} (F_I^\pm)^{\b\n}\ .}
Then we contract \quatrss\ by $G^{\n\b}$ and use \voith\ to obtain
\eqn\riic{ R_{\m\n} = -\ha R_{\m\a\b\g} (F_I^\pm)^{\b\g} (F_I^\pm)^\a{}_\n
\ - \ 2 \vec C^{\mp }_{\m\a}\cdot
(\del_{\vec \th}  F_I^\pm F_I^\pm)_{\n}{}^\a \ .}
This could also be given a different form using \thett. 
We see that generically
only when $N=4$ is local we obtain the usual Ricci flatness condition.
Next we specialize to an important case where the above conditions 
can be simplified considerably.

\bs
\no
$\underline {\rm A\ singlet\ and\ a\ doublet}$: Let us assume that 
there is only 
one $\th$. Having in mind the discussion of section 2 it can be easily seen 
that the only consistent possibility is to have two of the complex structures 
$F_{1,2}^\pm$ non--local and the third $F_3^\pm$ local. Without loss of 
generality we may choose
$F_2^\pm =\pm \del_\th F_1^\pm$ and then from \quaalg\ it 
follows that $F_1^\pm =\mp \del_\th F_2^\pm$. 
Thus $F_{1,2}^\pm$ form a doublet in each sector separately and $F_3^\pm$ is 
a singlet. 
Then a simple application of \thett\ gives  
\eqn\thed{ R^\pm_{\m\n\a\b} (F_J^\pm)^{\a\b} = \pm
d\ C^\mp_{\m\n}\ \e^{J 1 2}\ ,}
where $d$ is the dimension of the target space. 
We see that only for the singlet, corresponding to the local complex structure,
we get a non--zero result. For the doublet in contrast, we get zero.
Consistency conditions for \thed\ are obtained after
contracting it by $(F_3^\pm)^{\m\n}$ or by $(F_{1,2}^\pm)^{\m\n}$ and 
then using again \thed\ to reexpress the left hand side
\eqn\conth {(F_3^+)^{\m\n} C^+_{\m\n} + (F_3^-)^{\m\n} C^-_{\m\n} = 0\ ,
\qq (F_{1,2}^\pm)^{\m\n} C^{\pm}_{\m\n} = 0 \ .}

Before we turn to the torsionless case let us consider the case when the
non--locality arises from an abelian duality
transformation with respect to a Killing vector and, as usual, let us denote 
quantities in the dual model with tildes. Then, as we have mentioned,
$\tilde C^\pm_\m = \pm \tilde Q^\mp_{0\m}$ and hence 
$\tilde C^\pm_{\m\n} = \mp \ha  \del_{[\m} \tilde Q^\mp_{0\n]}$. 
Using the duality 
transformation rules for background fields \BUSCHER\ we derive that 
\eqn\cdual{ \tilde C^\pm_{\m\n} = ( A_\pm^T C^\pm A_\pm )_{\m\n} \ ,}
where $C^\pm_{\m\n} = \ha G_{00}^{-1} \del_{[\m} Q^{\mp}_{0\n]}$.
Then \thett, (written in the dual model with tildes) can be rewritten in terms 
of tensors defined in the original model, 
with the help of \tranc\curvte\cdual. We obtain 
\eqn\rtieine{ R^\pm_{\a\b\m\n} (F^\pm_J)^{\m\n} \ + \ G^{-1}_{00} 
\del_{[\a} Q^\pm_{0\b]} \del_\m Q^\mp_{0\n} (F^\pm_J)^{\m\n}
 \ = \ \ha G^{-1}_{00} 
\del_{[\a} Q^\pm_{0\b]} (\del_{\th} F^\pm_I F^\pm_K )_\g{}^\g \e^{IJK}\ ,}
where $I\neq J$. Assuming that the original model has manifest $N=4$ implies 
that the first term in the above equation is zero. If further 
the isometry is not chiral, i.e., $\del_\a Q^\pm_{0\b}\neq 0$, and using the
doublet structure of the two non--local complex structures mentioned above  
we deduce that  
\eqn\cdcdc{ \del_\m Q^\mp_{0\n} (F^\pm_J)^{\m\n} = \pm {d\ov 2}\ \e^{J 1 2}\ ,}
which notably, is only a dimension dependent constant. 
Notice also the consistency with \conth\
and that the left hand side of \cdcdc\ appears in \rdual.

If the torsion is zero then applying \riic\ for $I=1$ (or $I=2$) and using 
\thed\ gives 
\eqn\riiap{ R_{\m\n}= \mp  C^{\mp}_{(\m\a}\ (F_3)^{\a}{}_{\n)} \ .}
However, applying it for $I=3$ we obtain the same formula only if the target
space dimension is $d=4$. 
Therefore we have proved that the ansatz that
we can have non--local $N=4$ in torsionless backgrounds with one complex
structure being local and the other two forming a doublet is consistent
only in $d=4$ dimensions. In addition since the Ricci tensor $R_{\m\n}$ 
is the same in both chiral sectors we conclude that we should have 
$C^-_{\m\n} = - C^+_{\m\n}$ and therefore we may choose $C^-_\m = - C^+_\m$.

Using the explicit results of \basfep, we have verified 
the general structure we have exhibited here for 
hyper--kahler manifolds with a rotational Killing symmetry and their duals.
The same checking was also done using the results of section 4.

Lets us conclude this appendix by proving that we cannot have three non--local 
complex structures which 
depend on three functionals $\th^I$ and also transform as a triplet 
of $SO(3)$ in the sense that
\eqn\tratri{ \del_{\th^I} F_J^{\pm} = \e^{IJK} F_K^\pm\ .}
After some algebra we find that \thett\ takes the simplified form
\eqn\tripr{ R^\pm_{\m\n\a\b} (F_J^\pm)^{\a\b} = d\ C^{\mp J}_{\m\n}\ .}
However, since $C^{\mp J}_{\m\n}$ is independent of $\th^I$ we can take the 
derivative of both sides of \tripr\ with respect to $\th^I$, use \tratri\
to rewrite the left hand side and finally obtain that $R^\pm_{\m\n\a\b}
(F_J^\pm)^{\a\b}= 0$, which when compared to \tripr\ implies that
$C^{\pm J}_{\m\n} = 0$ for all three values of $J$. Therefore $C^{\pm J}_\m$ is
a total derivative and $\th^J$ is a local function.

\appendix B { ``Dynamical'' moduli and duality }

In this appendix we prove, for a class of models,
that making moduli parameters 
``dynamical'' (equivalently coordinate dependent) and retaining conformal
invariance is equivalent to performing duality transformations.

We start with an action corresponding to the tensor product
$G/U(1) \otimes (2D)_R$, where the second factor denotes any of the 
2-dim conformal models corresponding to the function $R(\r)$ introduced in 
section 4. Using the definitions \jjs\ we have
\eqn\rols{\eqalign{S= 
& I_0(h) +{1\ov 2\pi} \int E\ \del_+ \a \del_- \a
\ +\ {2\ov 1-\S} ( \del_+ \a J^-_i \del_- x^i + J^+_i \del_+ x^i \del_- \a +
 J^+_i J^-_j \del_+ x^i \del_- x^j )\cr
& + {1\ov 2\pi} \int \del_+ \r \del_- \r \ + \
 R^2(\r) \del_+\b \del_-\b \ ,\cr }}
where $E\equiv {1+\S\ov 1-\S}$. 
After we let $\a=\tau -\psi/2$ and $\b=\tau + \psi/2$ and perform a duality
transformation with respect to the Killing vector $\del/{\del\tau}$ we obtain 
the action 
\eqn\dualrol{\eqalign{ \tilde S= & I_0(h) + {1\ov 2\pi}\int \del_+ \r \del_- \r
+ {1\ov { E + R^2}} \biggl( \del_+ \tau \del_- \tau + E( R^2 \del_+ \psi
\del_-\psi + \del_+\psi \del_-\tau - \del_+ \tau \del_- \psi) \cr
& - 2 {1-R^2\ov 1-\S} J^+_i J^-_j \del_+ x^i \del_- x^j  \cr
 & +  {2\ov 1-\S} \bl( (\del_+ \tau - R^2 \del_+ \psi) J^-_i \del_- x^i 
- J^+_i \del_+ x^i (\del_-\tau +R^2 \del_- \psi) \br)\biggl) \ .\cr } } 
We would like to compare this action with the one that follows from the 
WZW model for a group $G$ marginally deformed by a current--bilinear in the
Cartan torus. This is equivalent to the model dual to \debo\ and the 
deformation parameter is the modulus $Q$. The idea is then, to make the modulus
a function of the target space variables \refs{\GSVYdyn,\KKdyn,\Tsdyn}.
In our case we add to the dual to \debo\ the term ${1\ov 2\pi} \int \del_+ 
\r \del_- \r$ and for convenience we let $\psi\to \ha (Q \psi + \tau)$. 
The result becomes just \dualrol\ if we replace $Q\to R(\r)$. 
Of course without trying to satisfy the $\b$-function 
equations we do not know a priori the function $R(\r)$. But the 
relation to the direct product $G/U(1) \otimes (2D)_R$ by a duality 
transformation tells us that conformal invariance constrains $R(\r)$ to be
given by \condR.

\listrefs
\end